\documentclass[aps,nofootinbib,notitlepage,longbibliography,twocolumn]{revtex4-1}
\usepackage{amsmath,amssymb}
\usepackage{slashed}
\usepackage{graphicx}
\usepackage{bm}
\usepackage[T1]{fontenc}
\usepackage[utf8]{inputenc}
\usepackage{gauss} 
\usepackage{ytableau}
\usepackage[top=1.0in,bottom=1.0in,left=1.0in,right=1.0in]{geometry}
\usepackage[colorlinks,linkcolor=blue,citecolor=blue]{hyperref}
\newcommand{\be}{\begin{equation}}
\newcommand{\ee}{\end{equation}}
\newcommand{\bea}{\begin{eqnarray}}
\newcommand{\eea}{\end{eqnarray}}
\newcommand{\ba}{\begin{eqnarray}}
\newcommand{\ea}{\end{eqnarray}}
\usepackage{graphicx} 

\begin{document}

\title{D-shell mixing in light baryons and its effect on the orbital motion}

\author{Nicholas Miesch}
\email{Nicholas.Miesch@stonybrook.edu}
\affiliation{Center for Nuclear Theory, Department of Physics and Astronomy, Stony Brook University, Stony Brook, New York 11794--3800, USA}

\author{Edward Shuryak}
\email{edward.shuryak@stonybrook.edu}
\affiliation{Center for Nuclear Theory, Department of Physics and Astronomy, Stony Brook University, Stony Brook, New York 11794--3800, USA}

\author{Ismail Zahed}
\email{ismail.zahed@stonybrook.edu}
\affiliation{Center for Nuclear Theory, Department of Physics and Astronomy, Stony Brook University, Stony Brook, New York 11794--3800, USA}

\begin{abstract} 
The standard description of the nucleon in the non-relativistic quark model is an $1S,L=0$ state without orbital motion.  Yet, there are several indications from phenomenology that an admixture 
of states with nonzero orbital motion
maybe substantial. In this paper we focus on
the ``second shell" of the nucleon excitations (D-shell), for which
we give a modern description of the wave functions. We follow it 
by investigating what we call a ``maximal mixing" scenario,
assuming a hypothetical long-range tensor force. 
 We give the explicit wave functions for all states, before and after mixing, and re-assess
many predictions such as the magnetic moments, the standard and transitional form-factors from the nucleon to $N^*$. 
Unexpectedly, in this scenario we can reproduce the
long-puzzling features of the Roper resonance $N^*( 1440)$.  But even in this extreme case,  the
 admixture of the $1D,L=2$ state to a nucleon remains signifiantly smaller
 than expected from phenomenology.
\end{abstract}

\maketitle

\section{Introduction}  
This paper is a continuation of our 
efforts to address some open issues in hadronic spectroscopy.
In our previous paper \cite{Miesch:2023hvl} we focused on the first shell (negative parity) $L=1$ baryon resonances. Yet, and besides some technical progress, we basically
confirmed the classic conclusion by \cite{Isgur:1978wd} that the
quasi-local spin-spin and tensor forces can describe the first shell states, 
while the spin-orbit forces need to be canceled
by a long-distance negative contribution following from the Thomas term due to the  QCD string.   

We then derived explicitly the  orbital-color-flavor-spin wave functions with Fermi statistics, using a novel method based on representations of the permutation group $S_3$. In the subsequent paper \cite{Miesch:2024vjk} we
generalized this approach to larger systems, with five, six and even 12 quarks. 

A natural step is an extension of this construction to the second shell (positive parity).
For that we define the basic antisymmetric states, and use them to evaluate various 
observables:  matrix elements (magnetic moments), transitional form-factors and so on. For a nucleon admixture to the tensor state, the main observables are  various form-factors.

As we hope to demonstrate, the  spectroscopy of  the second shell baryonic resonances is not
settled yet, with some open questions.
Large splittings between these states is challenging to perturbative (or in fact any short-range) spin-dependent potentials. 
One of 50-year-old puzzle for the quark model, is the large downward shift of the
famed Roper resonance $N^*(1440)$. Also the issue of 
 admixture amplitude of the nucleon to the second shell states,  remains  
unsettled. Using different
tensor potentials, we will explore the case of  ``a maximal mixing scenario",
induced by (so far hypothetical) a long-range tensor force.

Before outlining the main structure of this paper, we will make 
few more general remarks on the quark model and phenomenology. 
Our previous attempts to bridge the standard hadron spectroscopy
in the rest frame, with the light-cone description, are still 
facing open questions. 

One issue is the famous  ``nucleon spin problem":
 empirical structure functions (at low normalization scale)
 still suggest that  orbital motion contributes
 about half of the nucleon spin. Generally, the orbital motion can originate
 from  different  sources. One --investigated in this paper -- is
 via admixture to  the second shell states with $L=2$. It may
 also come via the nucleon admixture to $qqq+q\bar q$ pentaquark states. This mechanism, as well as other issues related to pentaquarks, is to be studied in our next paper.

This paper is structured as follows. We start with few short
introductory subsections, containing historic references, general information and, briefly,  the status of the issues to be discussed in the main part of the paper. In \ref{sec_second_shell} we remind the reader of the structure of the second shell baryons. 
In \ref{sec_ff_intro} we recall the main observables and puzzles associated with those. 
The nucleon spin problem is widely known and discussed, but for completeness
we will recall it briefly in section \ref{sec_nucleon_spin}, with emphasis on the orbital part.
Many technical details  -- e.g. explicit wave
functions --  are delegated to 
Appendices, while the main text focuses on the physics and the
calculation results.

\subsection{Spectroscopy of the second shell baryons}
\label{sec_second_shell}

 The positive parity excited states of the baryons have orbital momentum up to $L=2$ and spin up to $S=3/2$, so their total angular momentum can be $J=1/2..7/2$. In this paper we focus on light $u,d$ quark states.  Furthermore,
since we will be mostly interested in their admixture to the nucleon, we will focus on a subset of $N^*$ states with the same angular momentum and isospin, namely $I=1/2,J=1/2$. 

QCD-based quark models of baryons, including
constituent quark masses, confining potentials and perturbative spin-spin and tensor forces, have been developed in the 1970's.  Isgur and Karl in their classic papers, on the first \cite{Isgur:1978xj} and second \cite{Isgur:1978wd} shell, laid the foundation. Their model  described rather well
the spectrum  of the excited baryons known at that time,
and predicted many more which (nearly all) were found later.
For recent comprehensive reviews of the quark model of baryons see e.g. \cite{Capstick:2000qj} and~\cite{Gross:2022hyw}. 

By revisiting the baryon spectroscopy again, there is no need for
the  simplifying assumptions  made in those works.  
Instead of assuming Gaussian wave functions, one can
 calculate those for realistic confining forces
in the hyper-distance approximation. 
The wave function satisfying Fermi statistics can be
constructed explicitly, via representations of the permutation group.
For the latters the calculations can be carried 
using modern  analytic computer tools, such as Mathematica. 
The Spin-isospin
 space of ``monoms"  $2^6=64$ (for nonstrange baryons)
can be used  in spin-tensor form. All pertinent operators
of spin-dependent forces,  can be 
 simply represented as matrices in this space, via the
 ``KronekerProduct" command, giving a very straightforward shape to
 all calculations. In this work we will carry the mathematica and the
 standard group representation analyses for cross check.

The main motivation for the present work was to check 
explicitly whether the observed splittings of states in the second shell
yield  strong mixing, and in particular
derive some estimates of the  tensor mixing with
the ground state (proton).  In other papers of our series 
we derived the quark-quark central and spin-dependent forces using two ``vacuum structure" models,  the Instanton Liquid Model (ILM) originating from \cite{Shuryak:1981kj}, and more recently the
Dense Liquid Model including instanton-antiinstanton molecules \cite{Shuryak:2021fsu, Miesch:2023hvl}. Both models 
turns out to produce forces which are rather short-range,   good for their applications to quarkonia but not very useful for the second shell
baryons, with their WFs vanishing at small distances.
To address this issue, we will introduce certain hypothetical long-distance tensor forces,  which we will use to evaluate  some  observables, and  assess their phenomenological limitations.

\subsection{Magnetic moments and  form-factors} \label{sec_ff_intro}
The central observable related to hadronic spins is
their magnetic moments. And indeed, already in the 1960's 
it was shown that the magnetic moments are reasonably described  by a non-relativistic $qqq$ models
$without$ any orbital admixture. If admixture of $L\neq 0$ states, either
in $qqq$ or  pentaquark $qqqq\bar q$ sector is there, some
 mechanism 
of cancellations between the spin and orbital contributions in the 
composition of the magnetic moments be needed.

The next sensitive observables are ``transitional" matrix elements between  on-shell states of pertinent operators (e.g. electromagnetic currents) with a finite momentum transfer, or transitional form-factors.
Starting from the diagonal nucleon-nucleon form-factors,
these are the electric and  magnetic Sachs form-factors. The hard (photon)
current contains either quark charges (Dirac) or quark spins (Pauli).

For a while, it was accepted that the electric-to-magnetic  ratio of the
form-factors,  is independent of the  momentum transfer $Q$. This supported the idea that the spin and radial parts of the nucleon wave function may
factorize. Also, it was expected that the neutron electric  form-factor  $G^n_E$ be  zero.
And yet, experiments at JLAB \cite{Puckett:2011xg} have shown that 
both predictions are not supported by experiment. The ratio  
 $G^p_E/G^p_M$ decreases with $Q$, with even a possible sign flip 
 at the upper end of the available measured range or $Q^2\sim 8.5\, \rm GeV^2$. The electric form-factor of the neutron is nonzero $G^n_E\neq 0$. 
 
 (Let us  note  that  the
 JLAB data  fall in the range of  $Q^2\leq 9\, \rm GeV^2$, or $Q/2\approx  1.5 \,\rm GeV$. These experiments are neither at small
 $Q$ where a soft chiral cloud is expected to dominate, nor
 at large  $Q$ where hard perturbative predictions are expected to work. This
is the intermediate or ``semi-hard" region,
 where form-factors have to be calculated
 non-perturbatively, see discussion in \cite{Shuryak:2020ktq}.)
 
How the
mixing between states may affect the form-factors, has been discussed
in literature. Specifically, the mixing
of the ground state nucleons
(proton and neutron) with states of the second shell
was  addressed in
 \cite{Simonov:2020wql}.  
 In this work, some schematic model was shown to reproduce
  the JLAB  data on the electric and magnetic, proton and neutron form-factors. Yet for this to happen it requires a rather strong and specific mixing.
As the reader will see from our analysis  below, we disagree with the  selection of the states in this work, dominant in the admixture, and its mechanism.


When evaluating the form-factors, we encountered a number of issues
which we would like to mention here. At large momentum transfer $Q\gg M$,  relativistic effects are important, so it is natural  
to  boost wave functions, from the CM to the light-cone. Examples  for the nucleon and delta are given in~\cite{Shuryak:2022wtk}, where the form-factors are also analyzed.

At small $Q$, we may naively  use the
Fourier transforms of the non-relativistic CM wave functions,
as done e.g. in atomic physics. Yet it does not work: one reason is that constituent quarks ,
unlike electrons,  are not point-like objects. The radii of the baryons are affected by  their ``chiral clouds".   

Yet at intermediate $Q$ 
the  nonrelativistic wavefunctions  can still be used, provided a ``simplified boost"   (e.g. in \cite{Simonov:2020wql})
$$\psi(\vec p_\perp, p_{l}) \rightarrow  {1\over  \gamma}\psi(\vec p_\perp, p_{l}/\gamma) $$
is made, via  compression of the longitudinal sizes  by a relativistic gamma factor $1/\gamma=\sqrt{1-v^2}$ .
Below, we will discuss to what extent this prescription can
be used to described the form-factors in the intermediate ``semi-hard"
range.


Furthermore, the {\em transitional form-factors}
from the nucleon to $N^*$ and $\Delta$ resonances,
provide a large set of
observables, sensitive to their internal structure. In particular,  the  Roper resonance $N^* (1440)$ famously show a transitional form-factor   in clear disagreement  with 
a ``minimal scenario", ascribing it to the radial 2S-excitation.
This fact is another strong motivation for  the present work.

\subsection{The nucleon spin/orbital problem} \label{sec_nucleon_spin}
The spin structure of the nucleon is a widely discussed issue, which does not  need an extensive introduction.
There is no shortage of in-depth reviews, e.g. \cite{Deur:2018roz}. The so called spin sum rule
\be {1 \over 2}= {1 \over 2}\Delta\Sigma +L_q+L_g+\Delta G \ee
where the first (last) term is the integrated 
quark (gluon) Parton Distribution Functions  (PDFs) of a polarized nucleon, e.g.  
\be \Delta\Sigma=\sum_q \int dx \Delta q(x) 
\ee
The naive expectation that the first term is  dominant, was shattered by experimental data, and more recently  by first-principle lattice simulations. Only about $half$ 
of the nucleon spin is coming from the quark spins, while the other half must come from the other contributions. 

Looking at this issue in more detail, one should clarify its $Q^2$ dependence. At large $Q^2$ it is defined by
perturbative evolution, and in the  semi-hard regime by
the magnitude of the twist-3 power effects. (We expect to
complete a separate paper on this issue.)
Most specialized reviews on the spin issue,  use either the standard lattice
normalization point $\mu=2\, \rm GeV$, or even higher scale
related to the actual momentum transfers in experiments.
The lattice results suggest the three contributions  $\Delta\Sigma,L_q,L_g$ be similar, or about $\sim 1/3$ each.

 Yet our  series of papers devoted to hadronic spectroscopy describe it
  at a ``low scale" $\mu \leq 1/\rho\sim   0.6\, \rm GeV$ 
 where there are no perturbative gluons.  The only glue is semiclassical, in  form of ``instantons"
 or their bosonized zero-modes using chiral fields, $\vec \pi,\sigma$.
Furthermore, in this paper we do not discuss 
the contributions of the  sea quarks.  In this case,
the orbital contribution is about $L_q\approx 0.6$
 (see e.g. our paper \cite{Liu:2024rdm}).

In summary, if the quark Orbital Angular Momentum  $L_q$ 
provides a  significant contribution to the nucleon spin, it is
 in direct contradiction with the non-relativistic  quark models
where  nucleons are assigned to the S-shell ($L=0$) states. 

\section{Theory and Phenomenology of the second shell baryons} \label{sec_second_shell}
\subsection{Generalities } \label{sec_generalities}
The two pillars of hadronic spectroscopy are
two non-perturbative phenomena characterizing the QCD vacuum.  
The first is the spontaneous breaking of chiral $SU(N_f)_c$ symmetry, leading to moving  {\em constituent quarks} with  effective masses $\sim 350\, \rm MeV$ (much larger than the so called {\em Lagrangian or current}
masses proportional to the Higgs VEV). 
The second is {\em electric color confinement} producing a {\em confining potential}, linear at large distances $\sigma r$. Both are believed to be
induced by topological solitons: instantons for chiral breaking,
and magnetic monopoles for confinement (for
extensive pedagogical text see e.g. \cite{Shuryak:2021vnj} and references therein).

The non-relativistic quark model was formulated in the 1960's, with light mesons and baryons, and complemented in the 1970's by heavy quarkonia. 
During the last decades the field of hadronic spectroscopy has been revived,
with the discovery of multiquark states, mostly by B-factories, as well as LHCb and other LHC detectors.

Three quark systems are traditionally described using two Jacobi coordinates
$\vec \rho,\vec \lambda$ (or momenta), with either location of the center of mass (CM) (or total momentum) put to zero,
see the definitions in Appendix \ref{sec_app_coord}.
If all three quarks have the same masses (light $u,d$, or $sss$ or heavy $ccc,bbb$) then the ground state wave function is approximated by a spherically symmetric 6-d wave function $\psi(Y)$ depending on the radial ``hyperdistance" variable \be Y^2=\vec \rho^2+\vec \lambda^2=R^2 \ee

The hyper-spherical (or hyper-central) approximation
originates from Nuclear Physics in the 1960's, where it 
was successfully used for the description of few-body $A=3,4...$ nuclei. 
If quark masses are the same, the
kinetic energy is proportional to a pertinent Laplacian. 
While the binary potentials are not 6-d spherically symmetric, their non-spherical components were evaluated
and found to be small.  Solving only the 6d radial Schroedinger equation is much simpler  than the use
of alternative approaches (like the Faddeev equation or
the expansion in a large set of basis functions).
By now it is considered as a standard method, 
used by multiple authors, which we will also follow here.  We have
already used it for baryons of the first shell \cite{Miesch:2023hvl}.  The actual calculations for the radial wave functions of the second shell are described in Appendix \ref{sec_wfs0}.

 The classic papers \cite{Isgur:1978xj} have considered the negative parity baryons  from the P-shell ($L=1$), and found that using two of them and disregarding the
spin-orbit, they achieve a good description of the observed spectrum. Recently we repeated this analysis in~\cite{Miesch:2023hvl} using explicitly the asymmetric wave functions and modern data, and came to the same conclusion:
the matrix elements of the spin-orbit $V_{SL}$ from our fit, are zero inside the error bars. 

This paper focuses on the second, positive parity shell
nucleon resonances. It may appear strange, since 
the Isgur-Karl model in~\cite{Isgur:1978wd} has also discussed this shell,  
for a model with only spin-spin and tensor forces, where
only one common parameter -- basically the gluon exchange coupling $\alpha_s$ fitted from the $N-\Delta$ splitting was used. 
Here we will also ignore the spin-orbit forces, and focus
on the spin-spin and tensor forces.

Yet, as we will show below, there is a rather large room for improvement. 
Before we come to our calculation, let us add some
general comments on that. The spread of the second shell resonances is large: their masses spread between 1400 and 2100 MeV. If in the zeroth order (without spin-dependent forces) one starts from the oscillator states 
(as done by Isgur-Karl  subsequent works)  with degenerate masses, this spread is attributed to
spin forces. If these  matrix elements are indeed large, $\sim 300\, \rm MeV$, they should completely
mix the second shell states. 
Furthermore, assuming that the matrix element between the nucleon and
the ``tensor" ($L=2$) state  is of similar magnitude,  one would expect the nucleon to strongly participate in such a mixing.  These are effects we are going to study in detail below.

\subsection{The radial ``hyperspherical"  equation}
The explicit calculation is detailed in the Appendices. While
most of the analysis  is rather standard, we decided to
detail it, including the explicit wave functions of the basis states.

In Appendix A we define the Jacobi coordinates used, the hyperdistance $Y$
(we use $Y=R$ inter-changeably to remain consistent with the notations in our  earlier papers)  and the 6-dimensional Laplace-Beltrami operator. The latter
includes the four angles of two 3d vectors $\vec\rho,\vec\lambda$, 
complemented by the fifth angle $\chi$.
Its eigenvalues follow from the radial hyper-spherical
equation, with ``realistic" confinement, which is solved numerically using Mathematica. The procedure and resulting radial wave functions are given in Appendix~\ref{sec_wfs0}.

We will mainly focus on the $J^P=1/2^+$ nucleons, starting from the nucleon itself, the 2S state, and three more states of ``the second shell" referred to as ``scalar", "vector" and ``tensor" (for the details see the Appendix). The radial hyper-spherical equation 
(we will refer to it as ``the zeroth order results") gives their masses as
\be 
H_0 =\rm diag\big(1070, 1665, 1956, 1956, 2048 \big)\, (\rm MeV) \ee
The results are by no means a degenerate shell of the
oscillator basis, as assumed in many works. In fact this spectrum is much closer
 to the experimental ones (From Particle data tables)
\be \big( 938, 1440, 1710 ,1880, 2100\big)\, (\rm MeV) \ee
The largest deviation (about $-200\, \rm MeV$) is seen for the second state,   the famous ``Roper resonance puzzle" debated in the literature for 60 years since its discovery. For the other states, the 
shifts are smaller.  One may therefore ask which spin-related forces can mix these states in such fashion: we will provide an example of those below.
Other room for improvements compared to literature we found is that
non-Gaussian shape of the wave functions significantly  impact 
the resulting  form-factors.

\subsection{Orbital-spin-isospin structure of the wave functions}
For baryons the antisymmetric color wave functions factorizes, and so
the next step is to combine multiple parts of  the orbital-spin-isospin wave function into a unique combinations
consistent with Fermi Statistics (that is being symmetric under
quark permutations), see the next subsection. Given that, one 
then construct the matrix elements of the spin-dependent potentials, i.e. spin-spin, tensor and spin-orbit $V_{SS},V_{T},V_{SL}$ respectively.

The explicit form of the orbital-spin-flavor wave functions is 
rather  technical, and is summarized in the Appendices. Here we will formulate
an alternative to the traditional representations of the $SU(6)$ spin-flavor group, as was done by Isgur-Karl and all subsequent literature. (One may find a bit more detail on this construction in the Appendix of \cite{Capstick:2000qj}.
We have not found where they are given explicitly, to check both 
 their permutation symmetry and pertinent $J$ value.)

Before going further, let us discuss the possible mixing by spin-dependent forces. The spin-spin $V_{SS}$  does not change the orbital momentum,  so it will only have diagonal matrix elements.
The tensor $V_T$ contains and $L=2$ part, so it has nonzero matrix elements
between all the  $L=0$ and the  $\psi^C$ $L=2$ states, particularly
for the nucleon. (Our mixing matrix has many zeroes, which
we do not see in the reported Isgur-Karl mixing matrix. It is not puzzling
in general as the definition of states are different.)

Anyway, we will 
reconstruct these states with a  focus on Fermi statistics, 
in other words asymmetry under all quark permutations.
The orbital-flavor-spin wavefunctions for
the D-shell were  rederived in~\cite{Miesch:2023hvl} from the
representations of the $S_3$
symmetry group generators, deriving them explicitly as a  tensor product of four objects $$ coordinate\times coordinate\times flavor\times spin $$ 
For completeness and as a check, the group theoretical construction is also summarised in the  Appendices.

\begin{figure}
    \centering
    \includegraphics[width=0.85\linewidth]{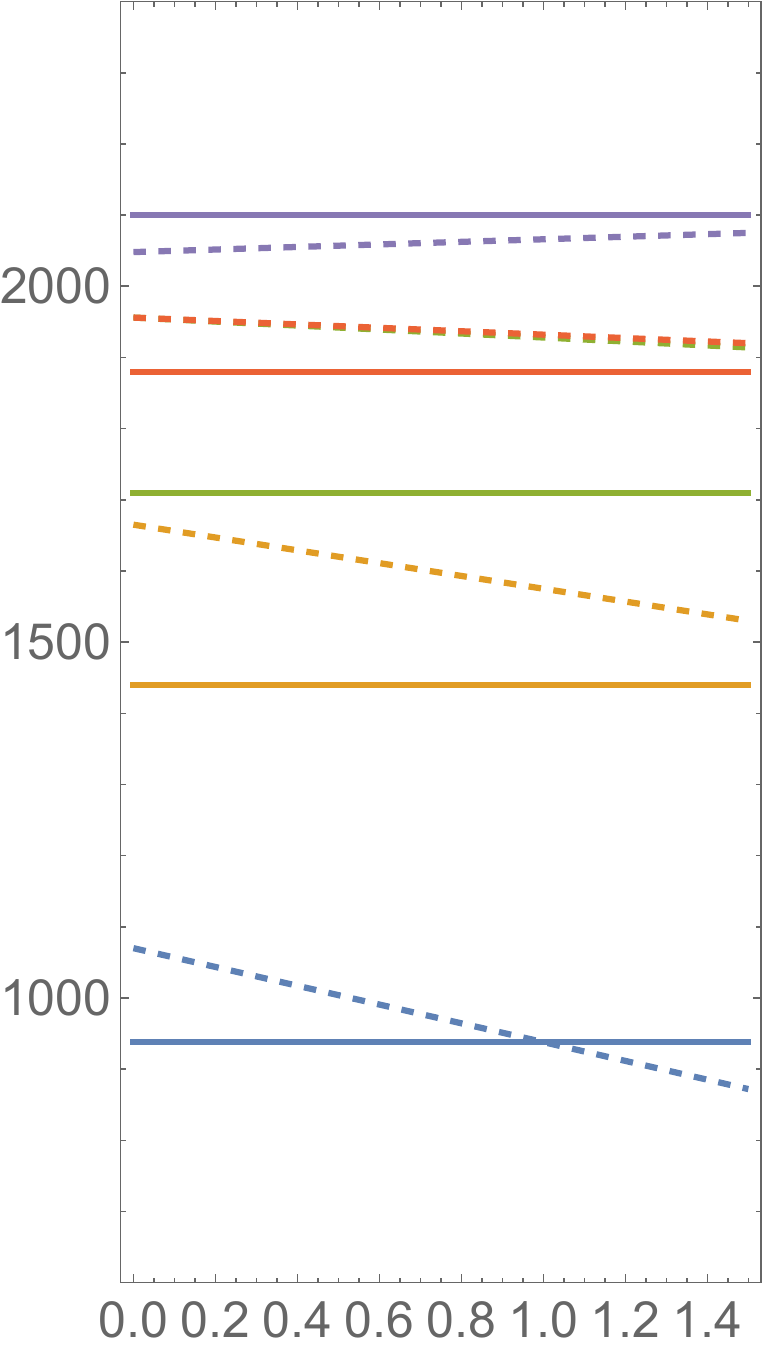}
    \caption{The five horizontal colored lines indicate the observed masses of the 
    nucleon and other four $J^P=1/2^+$ resonances, in MeV, from the Particle Data Tables.     The horizontal coordinates indicate the strength of the spin-spin forces, and the dashed line show the corresponding masses
    from the diagonalized Hamiltonian containing
    only $V_{SS}$.}
     \label{fig_masses1}
\end{figure}

\subsection{Spin-dependent forces}
The general expression for interaction of quark pair $i,j$ is used in the form
\bea
H_{sd}^{ij}=&&V_{SS}^{ij}\,\vec S_j\cdot \vec S_j\nonumber\\
&&+ V_{T}^{ij}\,\big[
(\vec S_i\cdot  \vec n_{ij})
(\vec S_j \cdot\vec n_{ij})- (\vec S_i \cdot\vec S_j)/3\big] \nonumber\\
\eea
where $\hat{ {\vec n}}_{ij}$ is the unit vector along the line connecting two quarks.  Following Isgur-Karl study and subsequent studies
(including ours), we neglect the spin-orbit contribution.
Since our wave functions are properly symmetrized under permutations, it is sufficient to calculate
their matrix elements for the $i=1,j=2$ pair, and then multiply the result by 3, the number of pairs. 

Using the explicit wave function described in the preceding section and the Appendix, we get the spin-dependent Hamiltonian as the following simple matrix
\begin{widetext}
\bea 
\label{eqn_Hsd}
H_{SS}=-\frac 34 \,\,
\begin{pmatrix}
  \langle V_{SS}^{1S1S}\rangle   & 0  & 0 & 0 & \\
  0  &   \langle V_{SS}^{2S2S}\rangle  & 0 & 0  & \\
     0   &  0 &  \langle V_{SS}^{3S3S}\rangle  & 0 & \\
         0   & 0  & 0 &  \langle V_{SS}^{1P1P}\rangle  & 0\\
            0    & 0  & 0 & 0 &  -\langle V_{SS}^{1D1D}\rangle  \\
\end{pmatrix} 
\eea
\end{widetext}
The first state is the nucleon (1S), the second is the (2S) state, and the remaining three are scalar (3S),vector (1P) and tensor (1D) states of the second shell, as defined above.
Although it is not shown explicitly, the matrix elements
involve coordinate-dependent
potentials, with the pertinent wave functions of the states.
Note that for 1-2 quark pair the inter-distance is $r_{12}=\sqrt{2} \rho $.
It  does not depend on the $\lambda$ coordinate, while the radial wave functions are 6d-spherical and depend on the hyperdistance $Y=R=\sqrt{\rho^2+\lambda^2}$.
Note that since the coefficient of $V_{SS}$ depends only on the total spin of the state, therefore only the last -- tensor $S=3/2$ -- state
has contribution with positive sign.

In Fig.\ref{fig_masses1} we show separately  the effects of  the spin-spin 
contributions  as a function of their strengths. The results follow from the diagonalization of the Hamiltonian $H_0+H_{SS}$, 
using the radial hyper-spherical equation described earlier.
The spin-spin forces  originate from both perturbative and non-perturbative effects, and have been studied on the lattice. Some details
are given in the  Appendix. Although their origins
are different, their contributions are dominated by
small distances, $r_{12}\sim 1 \,\rm GeV^{-1}$, and are 
stronger in the nucleon than in the second shell states.
If $V_{SS}$ is normalized to some magnitude such that at $x=1$ (the splitting gives the physical nucleon mass),
The near-local spin-spin forces cannot
describe the splittings in the second shell.

\subsection{ The tensor operator}
The pairwise tensor interactions are given by
\bea
\label{TTX}
\mathbb V_{T}(1,2,3)&=&\sum_{i<j=1}^3\,V_{T}^{ij} T^{i,j} \\
 T^{i,j} &=&(S_i\cdot \hat{r}_{ij}S_j\cdot \hat{r}_{ij} -S_i\cdot S_j/3)\nonumber\\
\eea
They split the even-parity  states of the second shell, and 
can induce  mixture with other shells, e.g. the nucleon. 
This operator can be averaged over five
orbit-spin-isospin wave function derived and given explicitly in Appendices.
Most of the values in the ensuing  5x5 matrix are zero, except when the "tensor" state is involved.

The 12 relative distance in the tensor  is $$r^{12}=\sqrt{2}\rho=Y\, \sqrt{2}\,\rm cos(\chi)$$ 
which means that the integration over four
angles (two solid angles) can be performed  explicitly
$$\int {d\phi_\rho\,d\theta_\rho\,d\phi_\lambda\,d\theta_\lambda \over (4\pi)^{2}}\,\rm sin(\theta_\rho)*sin(\theta_\lambda)... $$
with the results
\begin{widetext}
\ba
\langle P,1S | T^{12} (Y\,\rm cos(\chi)) |  VL2 \rangle_{\phi_\rho,\theta_\rho,\phi_\lambda,\theta_\lambda} &=&
\,\rm cos(\chi)^2/(3 \sqrt{5})  \\
\langle 2S | T^{12} (Y\,\rm cos(\chi)) |  VL2 \rangle_{\phi_\rho,\theta_\rho,\phi_\lambda,\theta_\lambda} &=& 
\,\rm cos(\chi)^2/(3 \sqrt{5}) \nonumber \\
\langle VL0 | T^{12} (Y\,\rm cos(\chi)) |  VL2 \rangle_{\phi_\rho,\theta_\rho,\phi_\lambda,\theta_\lambda} &=&
1/9 \sqrt{2/5}\,\rm  cos(\chi)^2 (2 - 5 cos(2 \chi)) \nonumber \\
\langle VL1 | T^{12} (Y\,\rm  cos(\chi)) |  VL2 \rangle_{\phi_\rho,\theta_\rho,\phi_\lambda,\theta_\lambda} &=&
\,\rm -{sin(2 \chi)^2\over 9 \sqrt{10}} \nonumber \\
\langle VL2 | T^{12} (Y\,\rm  cos(\chi)) |  VL2 \rangle_{\phi_\rho,\theta_\rho,\phi_\lambda,\theta_\lambda} &=&
(2/45)\,\rm  cos(\chi)^2 (-5 + 2 cos(2 \chi)) \nonumber 
\ea 
The right-hand-side still needs to be averaged over two remaining variables $Y,\chi$ with the appropriate radial
wave functions $\langle\psi^{I*}(Y)\, V_T^{12}(\chi)\,\psi^J(Y) \rangle $, with the $\chi$-measure
$$\int {d\chi \,\rm cos(\chi)^2 sin(\chi)^2 \over (\pi/16)}...$$
We will consider two limiting cases of the tensor forces, the perturbative ``short-range" tensor (second derivative of Coulomb potential)  and a hypothetical ``long-range" tensor
\be V_T^{\,\rm short\,range}\sim {1 \over \,\rm cos^3(\chi)},\,\,\, \,\,\,V_T^{\,\rm long\,range}\sim \,\rm cos(\chi)
\ee
to be motivated below. In the latter case the averaging over $\chi$ is straightforward with the result 
\ba
\langle P,1S | V_T^{12} | VL2 \rangle_{Y,\chi} &=&
 {128 \over 315 \sqrt{5} \pi} \langle P,1S | V_T^{12} | VL2 \rangle_{Y} \\
\langle 2S | V_T^{12}  | VL2 \rangle_{Y,\chi} &=& 
{128 \over 315 \sqrt{5} \pi }\langle 2S | V_T^{12}  | VL2 \rangle_{Y}\nonumber \\
\langle VL0 | V_T^{12}  | VL2 \rangle_{Y,\chi} &=&
{128 \sqrt{2/5} \over 2835 \pi} \langle VL0 | V_T^{12}  | VL2 \rangle_{Y} \nonumber \\
\langle VL1 | V_T^{12}| VL2 \rangle_{Y,\chi} &=&
  -{256 \sqrt{2/5}\over 2835 \pi} \langle VL1 | V_T^{12}| VL2 \rangle_{Y}  \nonumber \\
\langle VL2 | V_T^{12} | VL2 \rangle_{Y,\chi} &=&
-{3328 \over 14175 \pi} \langle VL2 | V_T^{12} | VL2 \rangle_{Y}  \nonumber 
\ea 
where in the right-hand side only averaging of the radial wave functions remains. The short-range case needs regularization, which  is discussed in Appendix.

\end{widetext}
The  tensor potential $V_T(r)$  can also be expressed as a
 correlator of Wilson lines appended by two magnetic field strengths \cite{Eichten:1980mw}. To our knowledge, it
has not yet been studied on the lattice. Forces due to  instantons or instanton molecules, by their nature, should only
apply at sufficient small distances $r_{12}\sim (1-2)\rho_{\,\rm inst}\sim (1-2)1/3 \,\rm fm$,  
 smaller than sizes of the second shell nucleon resonances in question. So, the potential $V_T$ at large distances remains mostly  unknown from first principles, besides phenomenological flavor dependent meson exchange models.

\subsection{Hypothetical long range forces}
Most of the studies of the tensor interaction in literature have assumed
its Coulomb-like perturbative form 
$V_T\sim \alpha_s/r^3$, and (inspite of the use of a 
large coupling constant $\alpha_s\sim O(1)$)
have concluded that the effects are small.
If we add the instanton-induced contribution, the results are still small
in light of their short range character. 
Yet, looking at the spin-orbit interactions, it was necessary to 
require that their short-range and {\rm long-range} contributions
cancel out in the first shell. If so, it is then natural to assume that in the second shell, the long-range effects are dominant.

\begin{figure}
    \centering
    \includegraphics[width=0.85\linewidth]{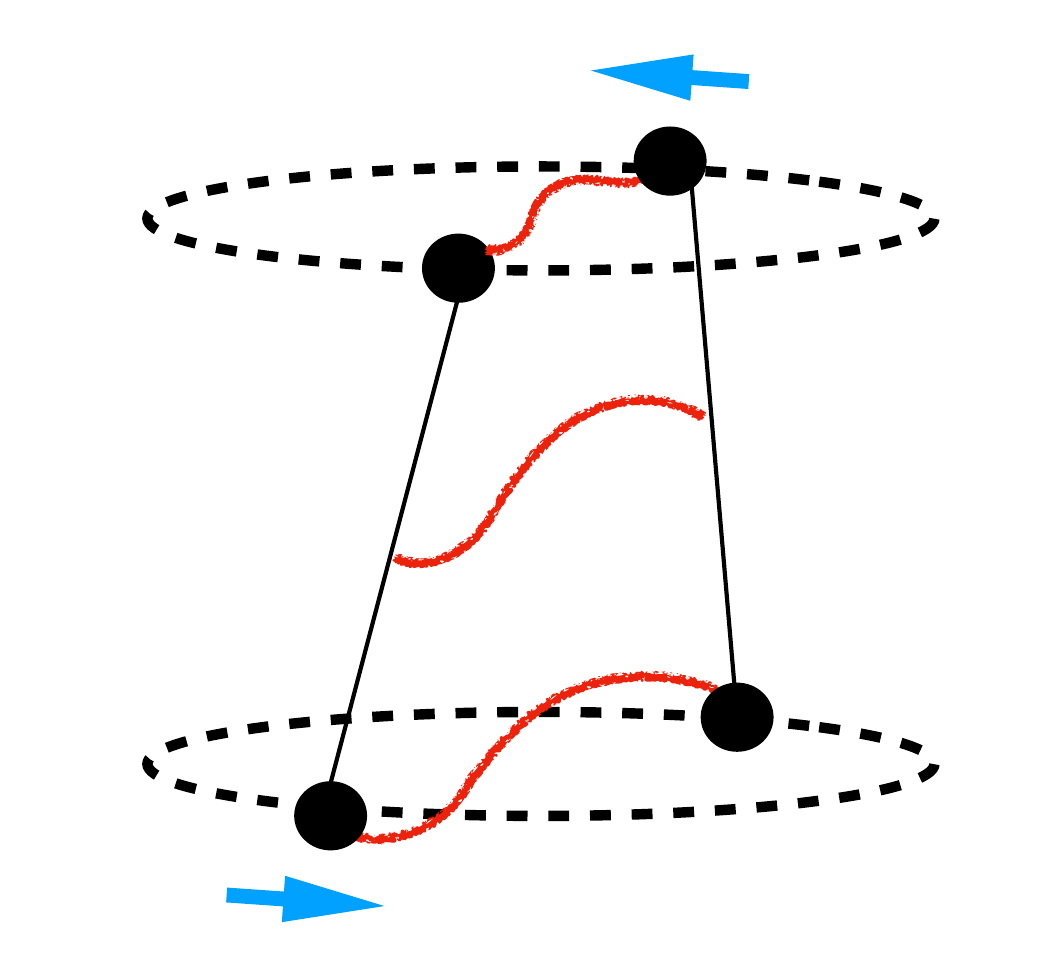}
    \caption{Sketch of two rotating quarks connected by a QCD string. Time is running upward, so
    a string is shown at three time slices. The quark spin is connected to the magnetic field related to the
    velocity-acceleration term, at the string ends.}
    \label{fig_rotating}
\end{figure}

A sketch of the long-range quark-quark interactions is shown in Fig.\ref{fig_rotating}. It is known that they should be dominated by
the QCD string. The black dots in the figure refer to certain operators 
describing the quark interactions with the string endpoints. 
Most studies describe the QCD string as a Nambu-Goto string, with
a simple scalar coupling to the quarks. However, 
generically, the end-points should carry both mass and spin couplings~\cite{Miesch:2024fhv}
\bea
\label{1}
S=&&\sigma_T\int_{-T/2}^{T/2}  d\tau\,d\sigma\, 
\sqrt{h}\nonumber\\
&&+\int_{-T/2}^{T/2}  d\tau\, \bigg(M\sqrt{{\dot{X}}^2}
-\frac 12 \sigma_{\mu\nu}
\frac{{\dot{X}}^\mu {\ddot{X}}^\nu}{{\dot{X}}^2}
\bigg)
\eea
with the volume element given by the determinant
\bea
h={\rm det}\bigg(\eta_{\mu\nu}\partial_aX^\mu\partial_bX^\nu\bigg)
\eea
The last boundary term is the spin factor which enforces the spin-orbit (Thomas precession-like) interaction at the end-point of the world lines~\cite{Strominger:1980xa,Polyakov:1988md}. Indeed, if the  string endpoints have  velocity and acceleration in different directions (e.g. if those are rotating in circles, as sketched in Fig.\ref{fig_rotating}) then spin-dependent forces arise. While the mechanism by which these forces are generated
is known, we suggest that a more accurate account for the quantum string motion 
should lead to additional spin-spin and tensor contributions. After all, the QCD string is thick with curvature corrections for intermediate lengths still unclear. 

\begin{figure}
        \includegraphics[width=0.85\linewidth]{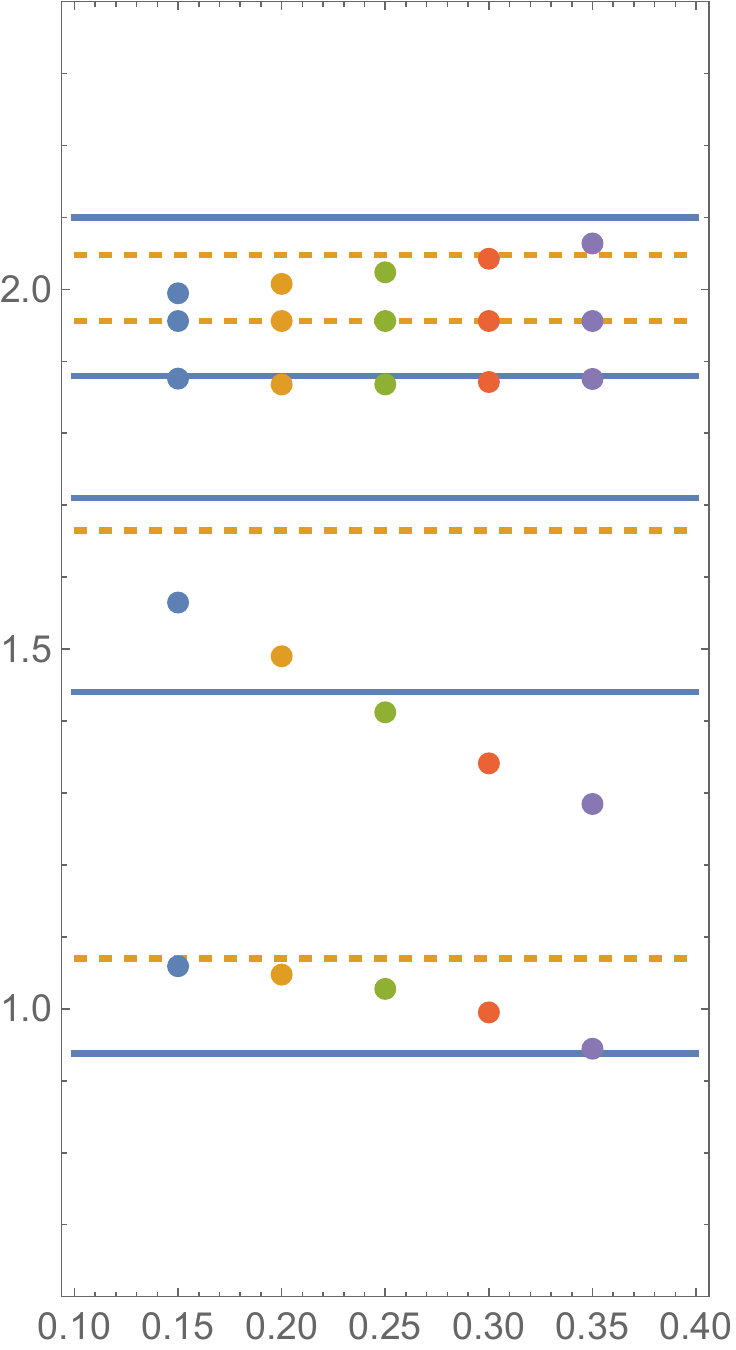}
    \caption{The horizontal solid lines are the observed masses of the 
    nucleon and 4 $J^P=1/2^+$ resonances from the Particle Data Tables in $\rm GeV$. The horizontal dashed lines the are masses calculated from the radial hyper-distance equation before mixing.
    The horizontal coordinate indicates the strength $\alpha_T$ of the hypothetical tensor potential
    defined in (\ref{eqn_tensor_longrange}). The points are the calculated masses by diagonalization of the Hamiltonian in the presence of the long rang tensor interaction. }
    \label{fig_masses2}
\end{figure}

Here we propose that the standard 
long range string confining potential $V_{conf}\sim r$
produced by the classical electric flux tube,  maybe extended to 
include spin-dependent coupling operators correlating the string end-points 
in a quantum string. In so far, we have not  worked out such operators, yet
we will provide a preliminary study of what such
operators may lead to.  The results will be compared  with
phenomenological observations. As an example, 
 let us consider what will happen with the second shell states,
 if the tensor potential  possesses a long-range component, say 
 \be 
 \label{eqn_tensor_longrange}
 V_T^{\,\rm long\,range}(r)= \alpha_T\,r 
 \ee
with some strenght $\alpha_T $. Note that (\ref{eqn_tensor_longrange}) is
comparable to the long range spin-orbit induced potential~\cite{Miesch:2024fhv} (and references therein), and much stronger than the string induced van-der-Walls potential in~\cite{Kogut:1981gm}.  
In Fig.~\ref{fig_masses2} we show the results of the lowest lying $J^P={\frac 12}^+$ nucleon resonances  versus $\alpha_T$.
The solid-lines are the masses evaluated using solely the hyper-distance equation without mixing, while the dots are the results induced by 
(\ref{eqn_tensor_longrange}) through mixing between the S-D shells versus 
$\alpha_T$. 
The suggested novel ``long-range tensor forces" indeed create significant shifts of the low lying masses by S-D mixing..

\subsection{ Roper ``collectivization" by the tensor force}
Using a hypothetical ``long range tensor potential"
we do indeed find a strong mixing and motion of the lowest lying $J^P={\frac 12}^+$ states as shown  in Fig.~\ref{fig_masses2}.The strongest shift is the downward motion of   the second state. Starting at $\alpha_T=0$
as a $2S$ hyper-radial state, it crosses the horizontal
line at 1440 MeV, corresponding to the PDG name of the so called ``Roper
resonance" discovered 60 years ago. (Note that modern pole mass
fit of it gives even   lower value, of 1370 MeV. We still call it 1440 following tradition.)

The history of the Roper state  have been covered e.g. in a separate review \cite{Burkert:2017djo}, where many models are discussed. Apart of 
  unexpectedly small mass, the Roper resonance has an unusual $N\rightarrow N^*(1440)$ form-factor, difficult to describe by quark-based models. 
 Nevertheless the authors of the review  still conclude that the Roper is basically a ``radial excitation" of the nucleon, or the $2S$ state.

 In our novel scenario with a long range
tensor force, the low mass of the  Roper
resonance is naturally explained by
  strong mixing in the S-D shell states.
In terms of the five basis states we use,
its wave function  at $\alpha_T=0.25$  is 
\be | {\bf R} > \approx 
\{ 0.28, -0.68, -0.06, 0.12, 0.66
\} \ee
So, the 2S contribution  as the 1D  ``tensor" contribution, about  50$\%$ each. Even the first state 1S state (nucleon) is represented by a significant probability  of about $0.28^2=7.8 \% $.

Our explanation of the unusual feature of the Roper resonance is 
thus a strong mixing in the second shell, induced by a hypothetical long range tensor interaction to be measured using lattice simulations.
This mixing is reminiscent of a phenomenon known in nuclear physics as
``a collectivization" of  a particular state, producing
``giant resonances". This idea was put forth in~\cite{Brown:1959zzb}. 
In short, consider an  $N\times N$ Hamiltonian of the schematic form 
\ba H=-\begin{tabular}{|cccc|}
1 & 1 & ... & 1   \\
1 & 1 & ... & 1   \\
...   & & &            \\
1 & 1 & ... & 1   \\ \end{tabular}
\ea
with all  matrix elements being the same. One of its eigenvalues is $-N$,
while all the others are zero. Its eigenvector  ${1 \over \sqrt{N}}(1,1,...1)$ is a coherent (``collective") mixture of all basis states. This description of the Roper as a collectivized state, is in line 
with the Skyrmion description of the Roper resonance as a breathingh mode of the solitonic pion field~\cite{Zahed:1984qv}.
While in our scenario the Roper is not yet quite "giant", 
the probability to remain in the original 2S state is about 1/3, practically the same as for other states in the mix. 
So, we suggest the Roper resembles a ``giant-like resonance" of the second shell, although with-not-so-large number of intermixed states $N\sim 3$.

\subsection{Transitional form-factors}
\begin{figure}
    \centering
    \includegraphics[width=0.85\linewidth]{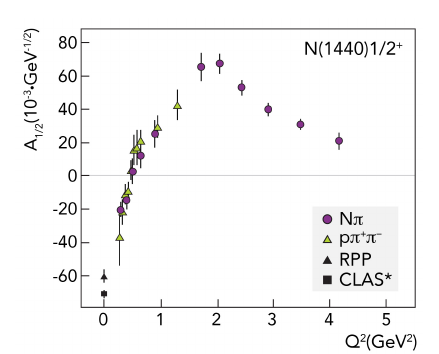}
        \includegraphics[width=0.75\linewidth]{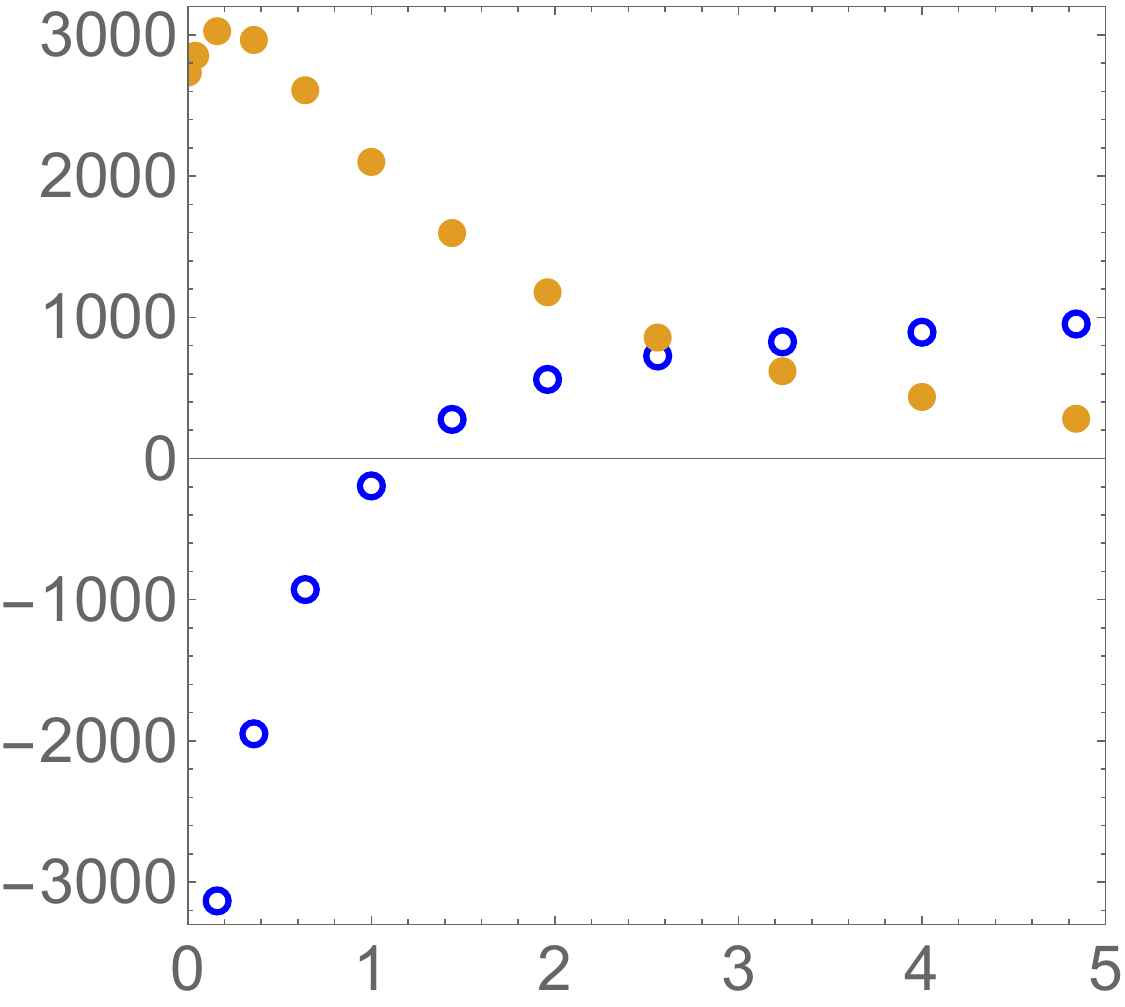}
    \caption{Upper plot shows experimental data on
 nucleon-roper transitional formfactor versus $Q^2 (GeV^2)$,  
.\\
 Lower plot show the  nucleon-roper transitional formfactor calculated for tensor coupling $\alpha_T=0$
  (open points) and $\alpha_T=0.25$  (closed points),
  corresponding to maximal mixing discussed in the text.}
    \label{fig_roper_FF}
\end{figure}

The transitions from the nucleon to other resonances
-- induced by the electromagnetic current --
have long history as well. They are the best way 
to check -- via their overlap -- how the resonance 
wave-functions differ from those for the nucleons. 
In Fig.~\ref{fig_roper_FF} (upper) (taken  from review \cite{Burkert:2017djo}) we show the current experimental situation for  the nucleon-Roper(N(1440)) transitional form-factor.
The lower plot shows our calculations for zero
and the maximal level mixing, using the  long-range tensor forces.
The mixing in the second shell changes
the shape of the form-factor drastically.
The  data is in between these two shapes, in support of the mixing. 

\subsection{Nucleon admixtures with the second shell  states}
The main physics message from
Fig.\ref{fig_masses2}, is that the mixing between the 1S nucleon state
and the second shell states, remains rather limited.  For the tensor coupling
$\alpha_T=0.25$ (which gives  correct Roper mass)
the wave function of the nucleon in our S-D basis, is
 \be \label{eqn_N_mixing}
 | N > \approx \{0.95, 0.11, 0.014, -0.029, -0.28
 \} \ee
The probability for the nucleon to remain in the original $1S$ state,
is  $90\%$, while the probability to be in the tensor second shell state  is $0.28^2\approx 7.8\%$. 
Such value should  be taken as an upper limit, corresponding
to a scenario with a ``good Roper". 
 
Could this  be the case in reality?
 Let us compare this result, with similar phenomena in nuclear physics. The most famous example is the
  deuteron, known to possess a finite quadrupole moment, due to a similar
  admixture between the  main S-wave (L=0) with the D-wave (L=2). 
  While we cannot compare the quadrupole moments, we can compare the admixture.
  The standard form of such an admixture is
  \be 
  \rm cos(\omega) \psi_{L=0} + sin(\omega) \psi_{L=2} 
  \ee
 which aounts to $\rm sin(\omega)\approx 0.2$, or $4\%$ in terms of the admixture probability. This admixture amplitude is only a factor of 2 smaller than our estimate for a nucleon.
 It is reasonable, since the tensor force is relativistic $O(v^2/c^2)$ effect, and the deuteron (with much smaller binding of only 2.2 MeV and 3 times larger
 mass of constituent $M_N$) is  less relativistic than the constituent quarks in the nucleon. 
  
A  recent study including the role of D-wave admixture in $\rm He^4$ has been carried out by one of us ~\cite{He:2024jgc}. In this study, the potentials without a tensor force were compared to those with a realistic Argonne v14 potential,  involving 14 operators in the  nuclear force. The admixture induced by the tensor operators was found to be in the  few percent range as well.

 To summarise: Even an assumed very strong long-range tensor force between
 constituent quarks, can produce only a moderate D-wave admixture with orbital momentum in the nucleon state. This admixture at the few percent level,  
is insufficient to explain several of the puzzles we mentioned in the Introduction.

\section{Magnetic moments}
The  magnetic moment operator in atomic physics has a long history,
going back to the definition $\vec \mu=2\vec S+\vec L$. While it is not directed along $\vec J $, its average is, with a cosine of the relative angle known as the so called Lande factor, a function of $ J,L,S $. 

The baryon  magnetic moments receive contributions from the intrinsic spin of each quark and their orbital motion,
\bea
\label{MUB}
\mu_X=\sum_{i=1}^3Q_i(g_S\,S^z_i+g_L\, L^z_i)\mu_Q
\eea
with $Q_i$ the ith quark flavor and $$\mu_Q=\frac {e\hbar}{2m_Q}=\frac{m_N}{m_Q}\mu_N$$ 
The ``consituent quark magnetic moment" is expressed in terms of the nuclear Bohr magneton, thus the mass ratio. For
spin-$\frac 12$ the Lande factors are $g_S=2g_L=2$. In terms of the Jacobi orbital angular momenta
$$\vec l_\rho=\vec \rho\times \vec p_\rho\qquad \vec l_\lambda=\vec \lambda\times \vec p_\lambda$$
and the ith-flavor magnetic moment $\mu_i=Q_i\mu_Q$.

Let us focus on  the 1-2 quark pair. The Jacobi coordinate $\vec \rho=(\vec r_1-\vec r_2)/\sqrt{2}$, and the orbital combinations 
$$\vec L_{1,2}=\vec r_{1,2} \times\frac 1i{\partial \over \partial \vec r_{1,2}}  $$  
 are the same, so each gives half of orbital momentum. Their contribution to the
 magnetic moment is
\be \mu_{12}^L={e_1\over 2e}L_1+{e_2\over 2e}L_2 \ee
where charges can be expressed via isospin Pauli matrices as $e_i/e=(1/6+\tau_3/2)$.

(\ref{MUB}) can be recast as 
\begin{widetext}
\bea
\label{MUB1}
\mu_X=g_S\sum_{i=1}^3\mu_iS^z_i+g_L\bigg(
\frac 12(\mu_1+\mu_2)l^z_\rho+\frac 16(\mu_1+\mu_2+4\mu_3)l^z_\lambda+
\frac 1{2\sqrt 3}(\mu_1-\mu_2)(\rho\times p_\lambda+\lambda\times p_\rho)^z\bigg)
\eea
\end{widetext}
Its average values for the 5 S-D basis states
denoted generically by $X$, normalized to that of the proton, are
\be {\mu_X\over \mu_p}=\bigg( 1,1,{1 \over 3},-{13 \over 108},-{49 \over 72}\bigg)
\ee
The signs follow the direction of the spin in all cases. Furthermore, even in the maximally mixed case discussed in this paper, the correction to the 
proton magnetic moment is only
$$-(49/72)\,0.078\approx -0.05$$
which is perhaps smaller than the current uncertainty of the  ``quark magneton" value $e/2M_q$. 


\section{Electric, magnetic and transitional formfactors } \label{sec_app_FF}
In the non-relativostic limit, the form factors are evaluated using
the wave functions. More specifically, in the commonly used Breit (brick-wall) frame, the elastic collision is  a transition between a hadron $i$ with momentum $\vec Q/2$ to a hadron $j$ with momentum $-\vec Q/2$ along 
the z-axis, caused by the action of the electromagnetic current operator,
$$\langle j| J_\mu^{em} | i \rangle $$ 
The first  step in the evaluation  of the form factors, is the calculation of
the Fourier transform of the wave functions, and the second step is their
convolution. As noted earlier, in the non-relativistic limit, we will use
the Lorentz contracted Fourier transforms
As we already mentioned in Introduction, as detailed in Appendix \ref{sec_app_FF}.

In Fig.\ref{fig_FF_1S1D}(upper) we show the form-factors 
calculated using  the 1S (blue dots) and 1D (orange dots) wave functions.
The comparison is to the empirical dipole fit
to the proton magnetic form-factor (dashed line). The 1D state is much softer 
than the 1S state as expected, since the 1S state is finite at the origin, 
while the  1D is quadratically suppressed by the centrifugal potential.

\begin{figure}
    \centering
    \includegraphics[width=0.75\linewidth]{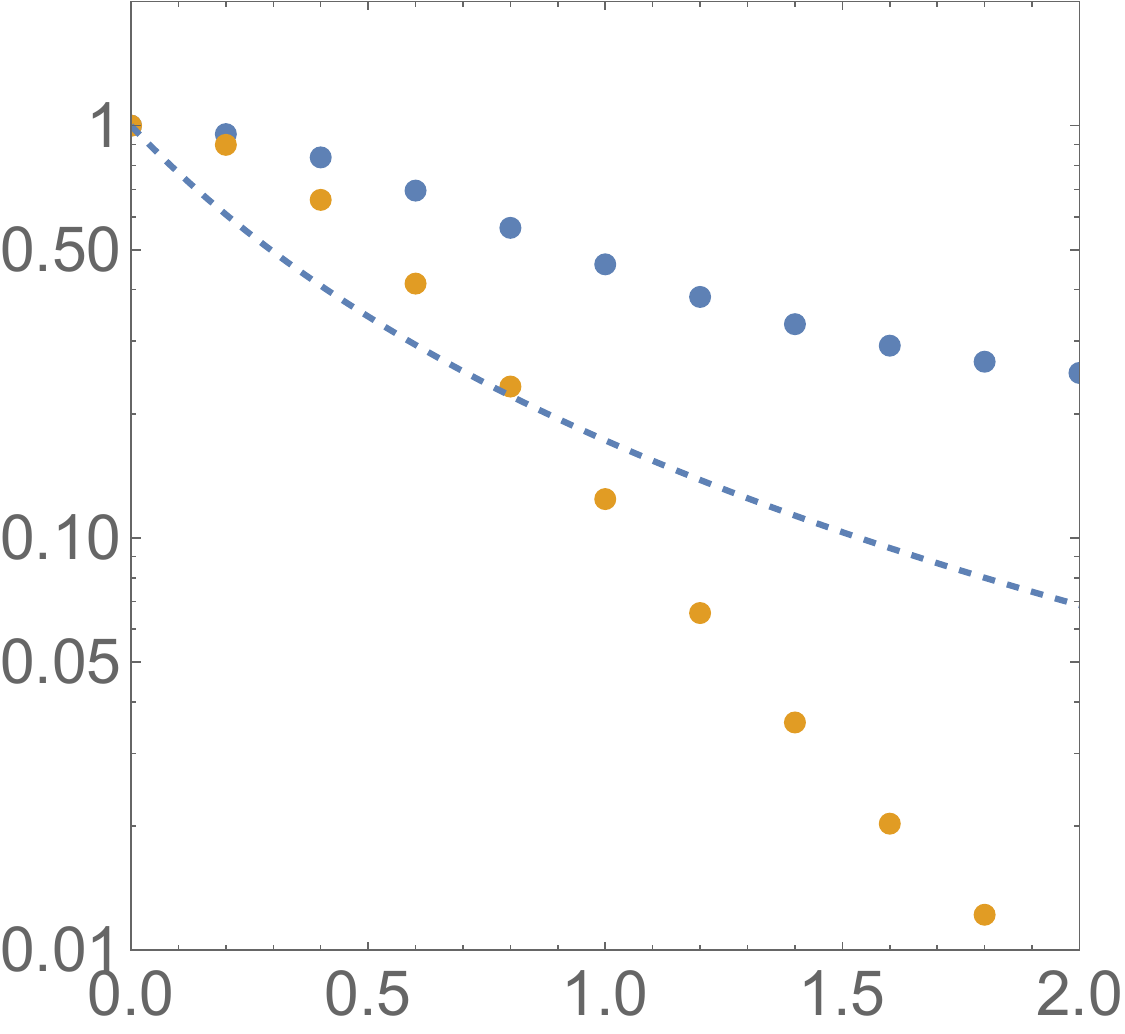}
        \includegraphics[width=0.75\linewidth]{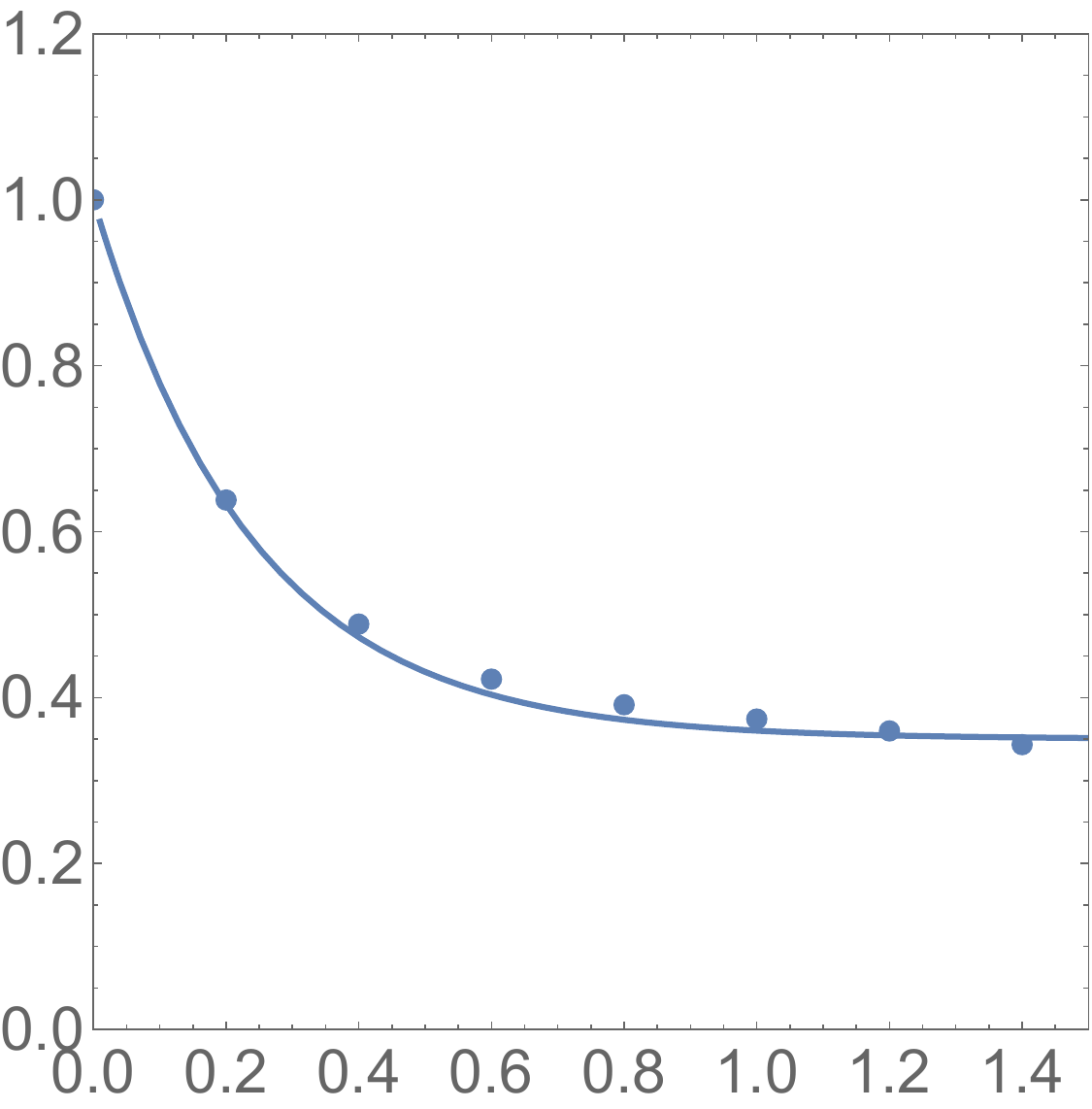}
    \caption{The upper plot shows the ${\frac 12}^+$  form factor using the 1S (blue dots) and 1D (orange dots) versus $Q^2 (\rm GeV^2)$. The comparison is to the nucleon dipole fit $FF_{dipole}=1/(1+Q^2/0.71)^2$ (dashed blue).\\
    The lower plot shows the ``constituent quark form-factor" versus $Q^2 (\rm GeV^2)$. The points are the ratio of the calculated form factor (points) to the fitted experimental curve (dashed) in the upper plot. The line is $0.35+0.65\,\rm exp(-Q^2R_{\rm cloud}^2/6)$ with radius $R_{\rm cloud}=1\,\rm fm\approx 5 \, \rm GeV^{-1}$. }
    \label{fig_FF_1S1D}
\end{figure}

Another important  feature is that the 1S form-factor
calculated from the hyper-central wavefunction 
(blue dots) does not  agree with the
data (the dashed line), even at small $Q^2$.
However, their behavior  at larger $Q^2>0.5 \,\rm GeV^2$ are in fact similar.
The fact that the hyper-central wave-function  strongly underestimates
the nucleon radius,  has been noticed by many authors and is well known,
 see e.g. \cite{Gallimore:2024fcz}. Indeed,
the standard radial wave-functions  are expected to give
the spatial distributions of the constituent quark $centers$, not the distributions of charge (or mass). The ``constituent quarks" are not point like objects but  have certain spatial distribution and form-factors by themselves.

To make this argument more quantitative, assume that the  experimental 
form-factor is the $product$
of two form factors, the the global wave-function described 
by the hyper-radius, times the constituent quark one, 
\be
\label{TIMES}
FF_{N}(Q^2)=FF_{\rm WF}(Q^2)\cdot FF_{\rm constituent\,quark}(Q^2) 
\ee
In the lower part of Fig.\ref{fig_FF_1S1D}, we show the last form factor
in (\ref{TIMES}). A point-like contribution from the constituent quark is present, which is about  $\sim 1/3$ of the total form factor.

At small momentum transfer it can be represented by a Gaussian $e^{-Q^2R^2/6}$ with a large radius $R\approx 5 \, \rm GeV^{-1}\approx 1\, fm$.
Such a form-factor is well described by the (instanton-based) semiclassical
theory of chiral symmetry breaking. To make a nonzero quark condensate, 
virtual quarks bound to an instanton (a zero mode) also have to reach the next (anti)instanton in an ensemble, located at distances $R\sim n_{\rm inst}^{1/4}\sim 1 \, \rm fm$. In other worlds, a constituent quark carries a  `` chiral cloud".

\begin{figure}
    \centering
    \includegraphics[width=0.85\linewidth]{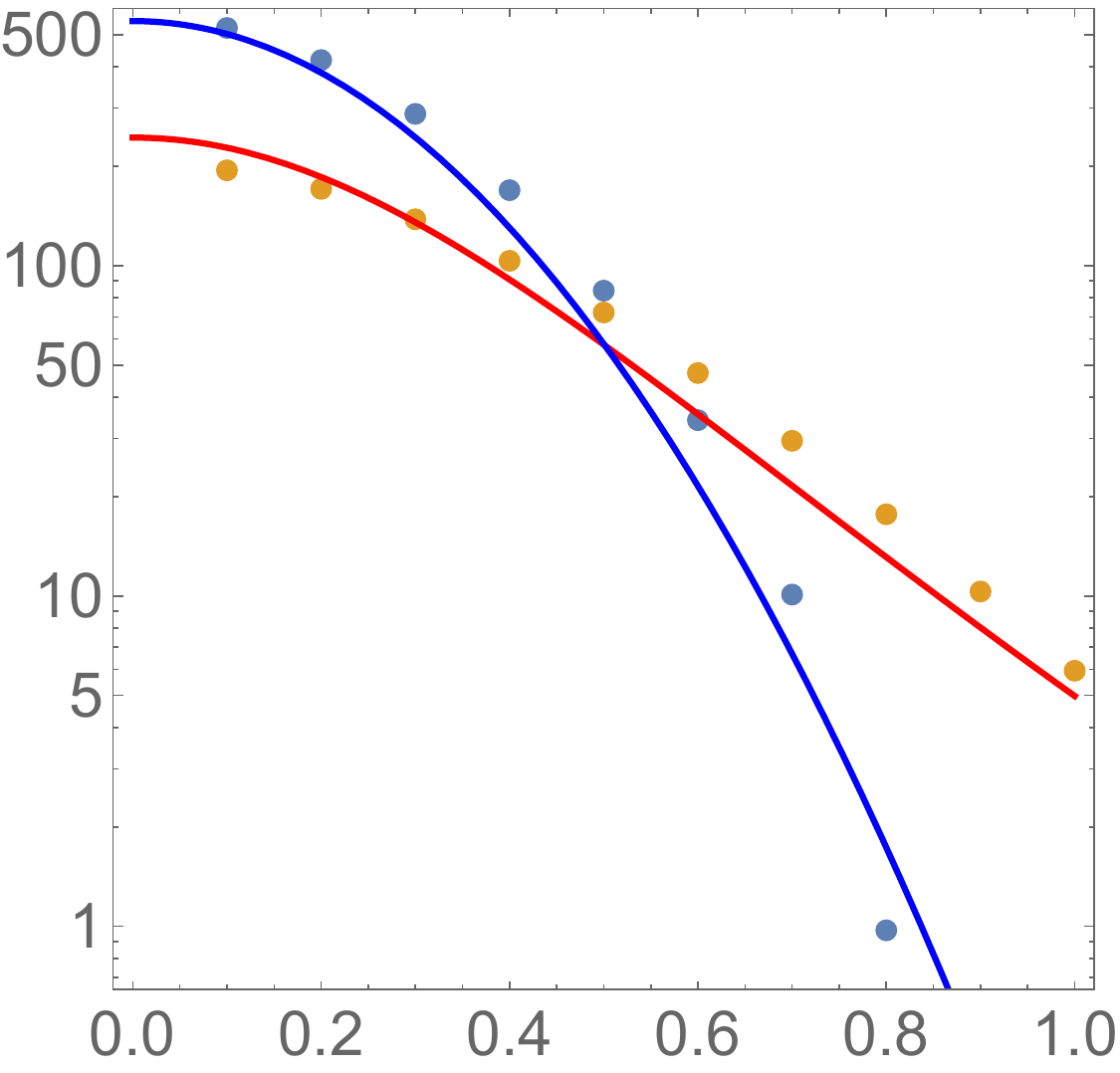}
    \caption{Red and blue points are for Fourier transform of the 1S and 1D radial wave functions
    versus momentum $p (\rm GeV)$. The solid lines 
    are the analytical fits in  (\ref{eqn_Fourier_app}).}
    \label{fig_Fourier_1S1D}
\end{figure}

In the remaining part of this section, we will briefly detail parts of the 
calculation entering the discussion above. More specifically, the  radial (hyper-distance) wave functions their 6-dimensional Fourier transforms 
 were all evaluated numerically. For analytical parametrizations  it is
 useful to recall the exact Fourier transforms for Gaussian and exponential
 wavefunctions
\ba  \rm exp\bigg(-{Y^2 \over 2 \Lambda^2} \bigg) &\rightarrow& \rm
exp\bigg(-{P^2 \Lambda^2\over 2 } \bigg) \nonumber \\
\rm exp\bigg(-Y\cdot A\bigg) &\rightarrow& {1 \over (1+P^2/A^2)^{7/2}}
\ea
With this in mind, the numerical form-factors  shown in~Fig.\ref{fig_Fourier_1S1D} for the 1S state  (blue dots)
and 1D state (orange dots) are well fitted by
\ba \label{eqn_Fourier_app}
\psi_{1S}(p) &\approx& {4 \pi 1.6 \over (0.7^2 + p^2)^{7/2}}\nonumber \\
\psi_{1D}(p) &\approx& 550.\cdot \rm exp(-9*p^2)
\ea
As noted earlier, the Fourier transform of the 1D state 
is softer than the 1S state, and not dominated by small distances
for this range of momenta.

The form-factor in the Breit frame is a convolution of the wave functions corresponding to total momenta $\vec Q/2$ and $-\vec Q/2$.
The boost amounts to dividing the longitudinal momenta 
by  the gamma factor $$1/\gamma=\sqrt{1-v^2}=\sqrt{1/(1+Q^2/4M^2)}$$ so the ``Lorentz-compressed" wave functions are $\phi=\psi(\vec p_\perp, p_{\rm long}/\gamma)$. The  form-factor includes scatterings on all three quarks, labeled  by $i=1,2,3$ and weighted by their charges $e_i$, 
\bea
\label{eqn_FF}
G_E(Q)=&&\sum_i {e_i \over e \gamma} \int {d^3 p_\rho d^3 p_\lambda}{(2\pi)^6}
\nonumber\\
&&\times\phi(p_\rho,p_\lambda) 
\phi(\vec p_\rho+\nu^i_\rho \vec Q,\vec p_\Lambda+\nu^i_\lambda \vec Q) \nonumber\\
\eea
for the proton $2/3,2/3,-1/3$, with 
 $\nu_\rho,\nu_\lambda$ referring to the weights of the respective Jacobi
 coordinates as fraction of the total momentum $\vec Q$. 

We now turn to the transitional form-factors, and the effect of level mixing.
Since in this work we have focused on 5 basis states,
we have calculated all $5\times 5$ transitional form-factors $\langle i |J| j \rangle$, where the in-out states consis of the "mixed" state following from
a direct diagonalization of the Hamiltonian. The chief  example, is  the transitional electromagnetic form-factor between the mixed nucleon state
{\bf N} and the  mixed  Roper state {\bf R}. Using the ``strong mixing" value $\alpha_T=0.25$ of the  tensor coupling, we obtain
\bea 
&&\langle{\bf N} | J |{\bf R} \rangle  \approx  -0.62 \langle 1 |J| 2 \rangle + 0.55 \langle 1 |J| 5\rangle\nonumber\\
&&+ 0.27 \langle 1| J | 1\rangle  + 0.26 \langle 2| J | 5 \rangle  - 0.18 \langle 5| J | 5 \rangle
\eea
Note that the original transition $ \langle 1 |J| 2 \rangle$ from 1S to 2S
state has the largest weight, but is not dominant. In fact the five contributions are comparable in strength
(additional terms with smaller coefficients  were dropped).
Since all the terms have different shapes, the configuration mixing
 leads to rather different shapes of the transitional form-factor. 
The results are shown in the lower part of Fig.~\ref{fig_roper_FF}.

\section{Summary and discussion}
The main questions discussed in this paper are: \\\\
(i) Can the observed splitting of the second shell baryons be understood by known (near-local) 
spin-dependent potentials? \\
(ii) Can a mixture of the nucleon with those states be quantified? \\
(iii) Can this mixing be reconciled with
other hints of a possible quark orbital motion inside the nucleon?
\\\\
The short answers to those questions are:\\
(i) No. The second shell baryons have large  sizes, so
the effects of near-local forces are small. In particular -- in agreement with  earlier
studies -- it is impossible to use them to explain  the puzzling properties of the Roper resonance.\\
(ii) A hypothetical long-range tensor force, does enhance the orbital  mixing of SD-states and provides a satisfactory description of the Roper resonance, as a  strong mixture of all second-shell states.  \\
(iii) Yet even this long range tensor force  does not affect the nucleon composition muh,
which remains dominantly the original 1S state. For the largest mixing, the ``tensor" amplittude is $\sim 0.28$, corresponding to a small probability $\sim 7.8\%$ in the nucleon. 
This mixing yields small corrections to the magnetic moments $\sim 5\%$. Both  changes should be regarded  as maximal, so the tensor mixing 
cannot explain the large orbital contribution reported from partonic data.\\

The traditional relativistic spin-dependent contributions (following
atomic and nuclear physics)  produce spin-spin, tensor and spin-orbit  potentials $V_{SS},V_T,V_{SL}$.  
 The perturbative one-gluon exchange gives local ($V_{SS}\sim \delta^3(r)$) or very short-range ($V_{SS}\sim 1/r^3$) potentials. The non-perturbative 
 and inhomogreneous field fluctuations we have studied in several of our previous papers, induced by instantons or  instanton-anti-instanton molecules, 
complement the perturbative results and improve the agreement with quarkonia
spectroscopy and lattice studies. Yet,
 by their very nature, they
are also rather short-range. We investigated these interactions using 
heavy quarkonia \cite{Miesch:2023hvl},  with overall agreement with earlier works.  More specifically, the $V_{SS},V_T$ are in agreement with phenomenological matrix elements,  except the spin-orbit $V_{SL}$ which  needs to be balanced by a long-distance negative contribution (tied to the Thomas contribution from the confining part). 
Yet, as we found in this work,
 all short-range effects do $not$  reproduce the splittings of the
second shell baryons.

While long range non-perturbative potentials from first principles are not
known (future lattice simulations would be welcome), we have posited the
possible existence of a ``long-range tensor force" perhaps  from a thick and confining quantum string. The result is a low mass  Roper resonance, and a tensor contribution to the nucleon state. This SD-mixing is found to be similar
in magnitude that observed in the deuteron and He-4 states in nuclear physics.

Finally, in this mixing scenario, we have analyzed the nucleon and the nucleon-to-Roper transitional form-factors. The description using  hyper-spherical wave functions  yields form-factors (and even r.m.s. nucleon radii) that are different from the data. While we have not analyzed this difference in detail,
we have shown that the difference can be explained by the fact  that hadrons are composed of constituent quarks,  which are $not$ just point-like, 
 but have intrinsic form-factors of their own. This situation is reminiscent of the low-energy electromagnetic form factors of nuclei, as composites of nucleons.

\newpage
{\centerline{\bf Acknowledgements}}
\vskip 0.5cm
This work is supported by the Office of Science, U.S. Department of Energy under Contract  No. DE-FG-88ER40388.
This research is also supported in part within the framework of the Quark-Gluon Tomography (QGT) Topical Collaboration, under contract no. DE-SC0023646.

\appendix 
\section{Coordinates} \label{sec_app_coord}
Throughout we made use of the Jacobi coordinates $\vec\rho,\vec\lambda$  traditionally used for baryons. They  are represented by a hyper-distance 
$Y=R$ and 5 angles  $\theta_\lambda, \phi_\lambda, \theta_\rho, \phi_\rho, \
\chi$ defined as follows 
\bea
x^a = Y&&\bigg( \rm cos(\chi) cos(\theta_\lambda), 
   cos(\chi) sin(\theta_\lambda) sin(\phi_\lambda), \nonumber\\
&&  \rm  cos(\chi) sin(\theta_\lambda) cos(\phi_\lambda) , 
  sin(\chi) cos(\theta_\rho) ,  \nonumber\\
&& \rm sin(\chi) sin(\theta_\rho) sin(\phi_\rho), 
   sin(\chi) sin(\theta_\rho) cos(\phi_\rho) \bigg) \nonumber\\
   \eea
with two 3-d solid angle variables  $\phi,\theta$, 
plus the 5-th angle $\lambda=Y {\rm cos}(\chi),\rho=Y \rm sin(\chi)$.
The volume element is
\be 
\sqrt{g}= Y^5 \rm cos(\chi)^2 sin(\theta_\lambda) sin(\theta_\rho) Sin[\
\chi]^2 
\ee
and the line element which defines the metric tensor, is 
\ba dl^2 &=& dY^2 + 
 Y^2 \rm \big(cos(\chi)^2 d\theta_\lambda^2 + d\chi^2  \nonumber \\
  &+&  \rm cos(\chi)^2 d\phi_\lambda^2 sin(\theta_\lambda)^2 + 
    d\theta_\rho^2 sin(\chi)^2   \nonumber \\
  &+&  \rm  d\phi_\rho^2 sin(\theta_\rho)^2 sin(\chi)^2\big) \ea
The corresponding Laplace-Beltrami operator is
\ba
 &-&{1 \over \sqrt{g}}{\partial\over \partial m} \sqrt{g} g^{mn} 
 {\partial\over \partial n }  \nonumber \\
&=&  {\partial^2 \over \partial Y^2}+{5 \over Y}{\partial \over \partial Y} \nonumber \\
&+&{1\over Y^2}\rm \bigg(sec(\chi)^2{\partial^2 \over \partial \theta_\lambda^2}  
+Csc[\theta_\lambda]^2 Sec[\chi]^2 {\partial^2 \over \partial \phi_\lambda^2} \nonumber \\
&+&\rm Csc[\chi]^2{\partial^2 \over \partial \theta_\rho^2} 
+Cot[\theta_\rho] (Csc[\chi]^2) {\partial \over \partial \theta_\rho} \nonumber \\
&+&\rm Csc[\theta_\rho]^2 Csc[\chi]^2{\partial^2 \over \partial \phi_\rho^2} \nonumber \\
&+&2 \rm Cos[2 \chi] Csc[\chi] Sec[\chi] {\partial \over \partial \chi} 
+{\partial^2 \over \partial \chi^2}\bigg)
\ea    

  \begin{figure}[t]
        \centering
               \includegraphics[width=0.85\linewidth]{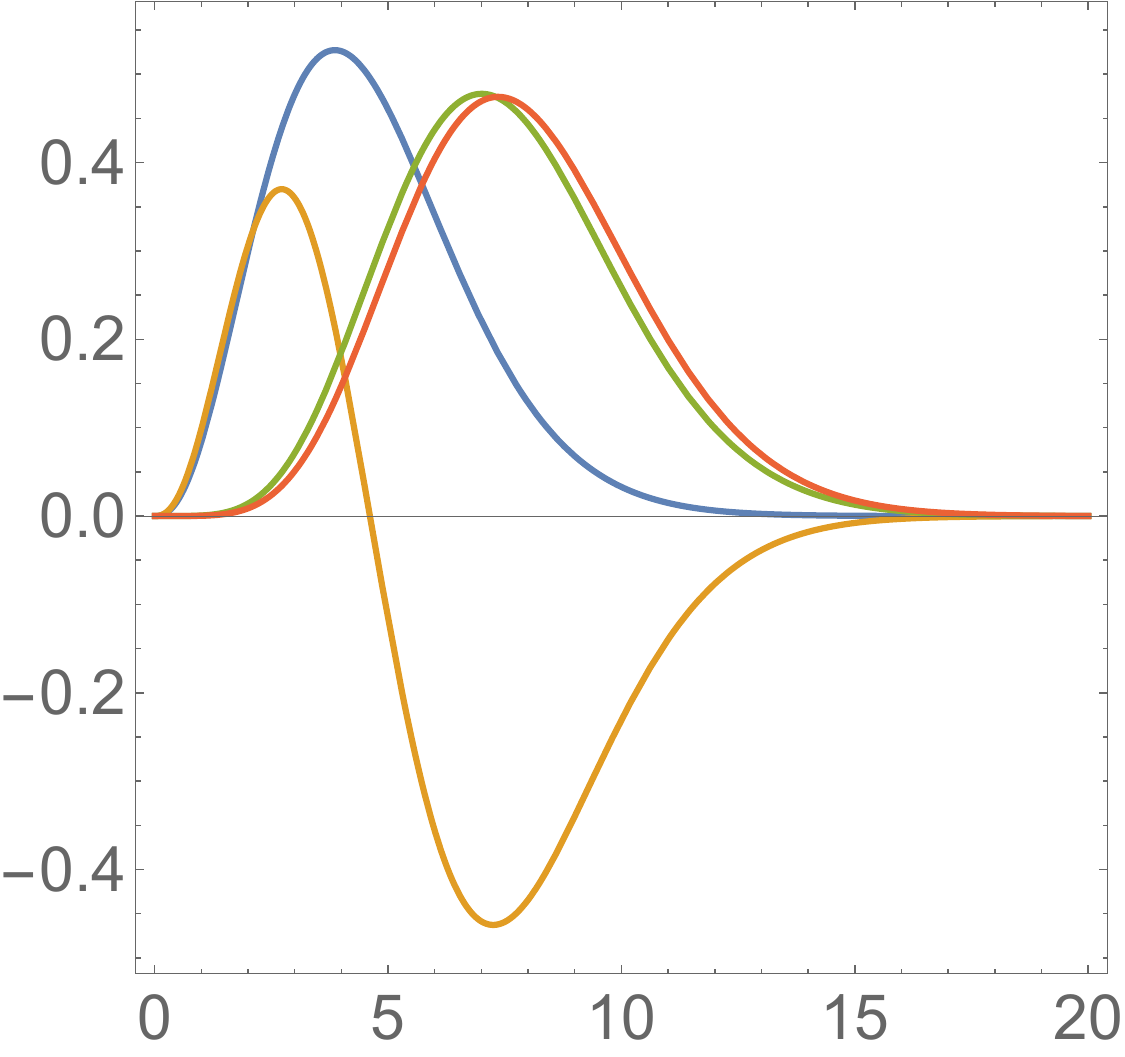}
        \includegraphics[width=0.85\linewidth]{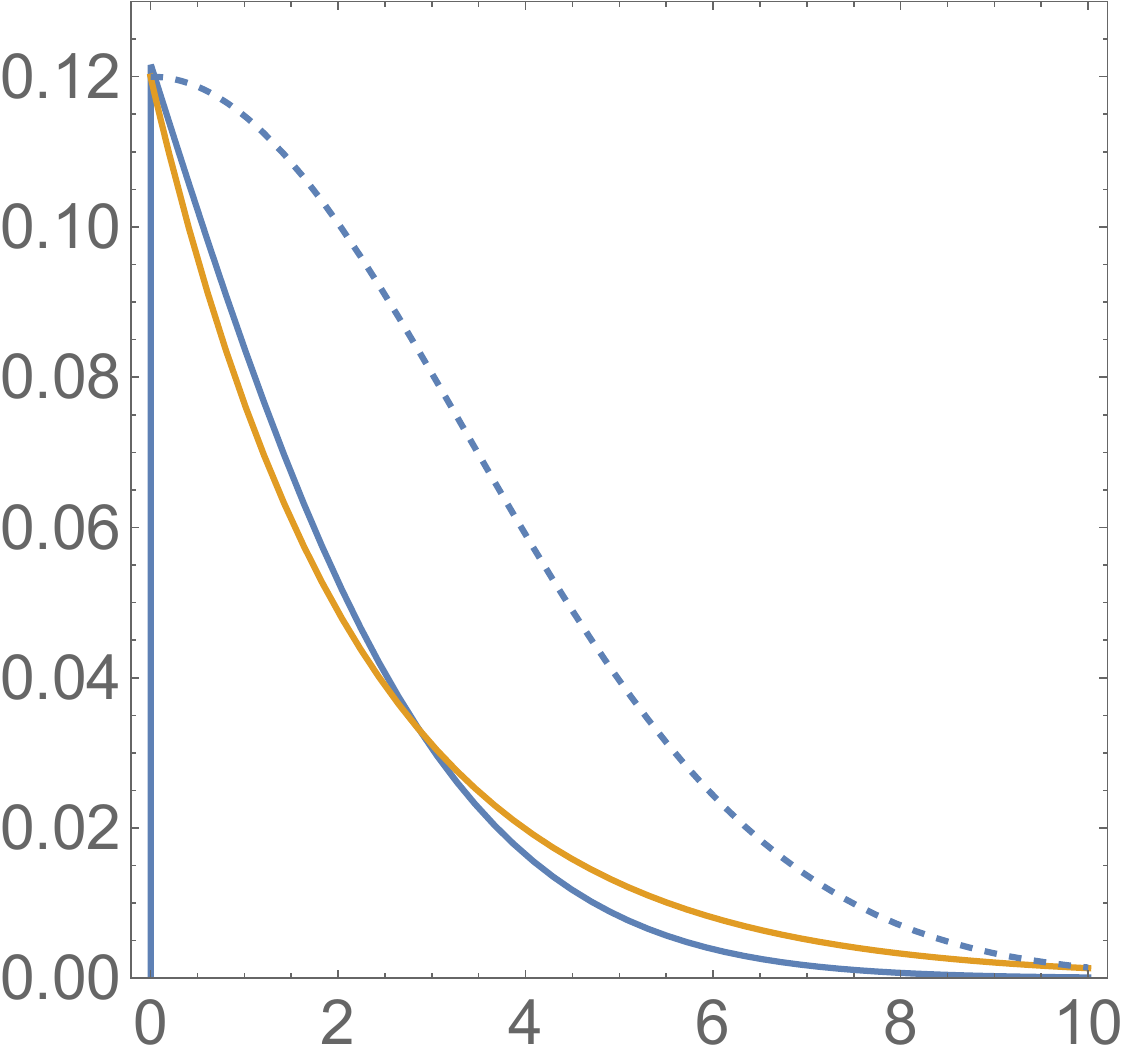}
        \caption{Upper: Hyper-spherical wave functions plotted as 
        $\psi(Y)Y^{5/2}$ versus the hyper-distance
        $Y\, (\rm GeV^{-1})$. The blue and orange curves are the $1S,2S$ states respectively, with zero angular Laplacian.  The green  and red curves correspond to the values  16 and 19, 
        respectively.\\
       Lower: The solid blue curve is the same numerical solution for the 1S state as in the upper figure, but now showing the  $1S$ wave function $\psi(Y)$ only. It is finite at the origin and about  exponential for  small $Y$. The red curve is the exponential approximation mentioned in the text.      
        The dashed curve shows the Gaussian approximation used by \cite{Simonov:2020wql}. All are normalized to the same wave function  value at the origin. }
        \label{fig_baryon_psi}
    \end{figure}

\section{Hyper-spherical wave functions} \label{sec_wfs0}
The hyper-spherical radial Schroedinger equation,  differs from
the usual one only by: (i) a 6-dimensional Laplacian in the kinetic energy, and (ii) a 6-d solid-angle average binary interaction potential
\be 
H \psi(Y)=-{1 \over 2M_q}
\bigg(\psi(Y)" +{5\over Y} \psi(Y)'\bigg) +(3M_q+\bar V) \psi(Y) 
\ee
with 
\be 
\bar V = 3\bigg( {32. \sqrt{2}\, Y \, 0.113 \over 
    30 \pi})\bigg)
    - {16. \sqrt{2}\over 3 \pi Y}\,\frac{0.64}2
    \ee
Here $Y=R$ is the hyper-distance and  $M_q = 0.35\, \rm GeV$ the constituent mass. We solve this equation for the $1S, 2S, 1D$ states  as shown in Fig.\ref{fig_baryon_psi}, with the normalization
\be \int |\psi(Y)|^2 Y^5dY =1\ee

The eigenvalues are the  masses in the lowest approximation (without spin-dependent forces). If the 1S mass is taken to be the spin-average $N,\Delta$
mass, then the corresponding three masses are 
\ba 
M_{1S}^0=1.07 \, \rm GeV\nonumber \\
M_{2S}^0=1.67 \, \rm GeV \nonumber \\ 
M_{1D}^0=1.82 \, \rm GeV 
\ea
Note that the last two numbers would be different in the oscillator model
used in \cite{Isgur:1978xj} and many subsequent works. Their difference $\sim 150\, MeV$ is not negligible and comparable to shifts due to spin-spin and tensor forces.

In the transition form-factor,  the dominant contribution stems from small sizes. In Fig.\ref{fig_baryon_psi} we  have chosen to
normalize the wave functions  to the same value at the origin. Out of the two approximations -- Gaussian used
in \cite{Simonov:2020wql} and our exponential fit -- we see that the latter is much closer to the numerical solution of the radial equation. (This is
to be expected, since the potential at small distances are
Coulomb-like rather than oscillator-like.)

\section{Symmetric orbital-spin-flavor WFs from representation of the permutation group $S_3$}  \label{sec_app_permuts}
The key constraint on the construction of the nucleon wave-functions is Fermi
statistics. The full wave-function is antisymmetric under any pair exchange
$P_{12}$, which amounts to 
$\rho \rightarrow -\rho,\lambda\rightarrow \lambda$,
hence the first angular combination is odd and the other even. Therefore
the flavor-spin WFs for the $\rho,\lambda$ contributions  must be different. 

The explicit  construction of the $\rho,\lambda$ parts of the wave-functions,
can be done in two ways. One, developed 
 in our previous paper \cite{Miesch:2023hvl}, is based on the representation of the permutation group $S_3$. Here
 we only need to add few things relevant for the second shell. The second derivation based on Young tableaux is discussed in the next Appendix. Both
 derivations were developed  for consistency checks.

We now recall the main ideas. The construction of the spin and isospin combinations
of the same form as the $\rho,\lambda$ coordinates, allows the building of the representation of the permutation group $S_3$, e.g.
\ba \label{eqn_Srho}
S_\rho &=& (\uparrow \downarrow \uparrow -\downarrow \uparrow \uparrow)/\sqrt{2} \\
S_\lambda &=& (\downarrow \uparrow \uparrow +
 \uparrow \downarrow \uparrow-2\uparrow \uparrow\downarrow)/\sqrt{6}
\nonumber 
\ea
The isospin states  $I_\rho,I_\lambda$ 
(or equivalently $F_\rho, F_\lambda$ as used in our previous work and the Young Tableaux below) are defined similarly.

Note that the permutation $P_{12}$ changes the sign of 
$S_\rho,I_\rho$ but not $S_\lambda,I_\lambda$, e.g. 
\bea
\label{eqn_perm_M12}
[P_2=(12)]
\begin{pmatrix}
M^\rho\\
M^\lambda
\end{pmatrix}
&=&
\begin{pmatrix}
-1&0\\
0 & 1
\end{pmatrix}
\begin{pmatrix}
M^\rho\\
M^\lambda
\end{pmatrix}\nonumber\\
{[P_4=(23)]}
\begin{pmatrix}
M^\rho\\
M^\lambda
\end{pmatrix}
&=&
\begin{pmatrix}
\frac 12&\frac{\sqrt 3}2\\
\frac{\sqrt{3}}2& -\frac 12
\end{pmatrix}
\begin{pmatrix}
M^\rho\\
M^\lambda
\end{pmatrix}
\eea
which clearly shows that the combination $\big(S_\rho I_\rho+ 
S_\lambda I_\lambda \big)$ is permutation invariant. 

In general, we will consider
a number of products of $N$ such objects, and will seek to find the
representation of the $N$-th tensor power of the $\rho,\lambda$  mixing matrices under permutations.

The cases discussed in this paper start with a nucleon in the ground state, 
with spin and isospin terms or $N=2$. In this case, we only have the symmetric combination  $S_\rho I_\rho+ S_\lambda I_\lambda$, as is well known.
In the first (negative parity) shell case $N=3$, as the coordinates are added
to the mix. In this case, there is only one symmetric combination.
\be V_{ABC}= A_\rho (B_\rho C_\lambda +  B_\lambda C_\rho) + A_\lambda (B_\rho C_\rho -  B_\lambda C_\lambda) 
\ee

The second shell has two coordinates, spin and isospin so in general it
requires four objects $N=4$, or the
construction of all  combinations  $A_\alpha B_\beta C_\gamma D_\delta$
with binary indices  $\rho,\lambda$. Those should be symmetric under any quark permutation. Selecting the permutations $1\leftrightarrow 2$ and $2\leftrightarrow 3$, as described by the  16-by-16 dimensional 
matrices
$$P_{12}^4=P_{12}\otimes P_{12}\otimes P_{12}\otimes P_{12}$$
$$P_{23}^4=P_{23}\otimes P_{23}\otimes P_{23}\otimes P_{23}$$
and diagonalizing them,  yield the number of 
symmetric and antisymmetric combinations. It turns out there are three
such combinations as given in (A11)  of \cite{Miesch:2023hvl},
\ba V_1 &=& A_\lambda B_\lambda C_\lambda D_\lambda \\
&+& A_\rho B_\lambda C_\lambda D_\rho+A_\lambda B_\rho C_\rho D_\lambda+A_\rho B_\rho C_\rho D_\rho \nonumber
\ea
\ba V_2 &=& A_\lambda B_\lambda C_\rho D_\rho \\
&-& A_\rho B_\lambda C_\lambda D_\rho-A_\lambda B_\rho C_\rho D_\lambda+A_\rho B_\rho C_\lambda D_\lambda \nonumber
\ea
\ba V_3 &=& (A_\rho B_\lambda - A_\lambda B_\rho)(C_\rho D_\lambda -C_\lambda D_\rho)
\ea

In total, we discuss 5 nucleon states, all with $J=1/2,I=1/2$:
the (1S)  nucleon,
the second (2S) state, two scalar $L=0$ states formed from the $V_1,V_2$ combinations defined above, the ``vector" $L=1$ formed from $V_3$ (including the vector product of $[\vec \rho \times \vec \lambda] $), and finally the ``tensor" $L=2,S=3/2$ combination. 
Using the standard Clebsch-Gordon coefficients, we first combine two coordinates into states with fixed $L=0,1,2$, and then combine them with
spin $S=1/2$ or $S=3/2$, as appropriate to get states with the desired $J=1/2$.
We use the spin-tensors in the spin-isospin monom space (64-dimensional) with
analytic manipulations  using Mathematica. The expressions in terms of monomons  are  too long to be listed here. 
They are orthogonal to each other and normalized to 1, after integration over all angles. Acting on them by the angular part of the Laplacian given above, we 
find that the  eigenvalues for those 6 states are
$ \big[ 0, 0, 0, 16, 19\big]$, respectively. 
Acting on them by spin-dependent forces lead to the matrices discussed in the main text.

\section{Alternative derivation of the orbital-spin-flavor WFs}  
\label{sec_permutations}



In this section we present the alternative derivation based on Young Tableaux.
The idea used already in~\cite{Isgur:1978xj},  is to construct mixed symmetry wave functions $M^{\rho, \lambda}$, out of Jacobi $\rho$-like and $\lambda$-like blocks which have pure S or A permutation properties,  both for spin and isospin parts. Throughout this appendix, we will adhere to
the notations used in~\cite{Isgur:1978xj} and also followed in our recent analysis of the P-states~\cite{Miesch:2023hvl}. More specifically, for the hyperdistance $Y\rightarrow R$, and
the integration measure will be given in terms of $d\vec\rho\, d\vec\lambda$ without carrying explicitly the
relative angular integration over $\chi$. Relative signs in transition matrix elements reflect on the
overall sign undermination of the wavefunction.

Construction of representations of $S_3$ is done like for any other groups, e.g. familiar
generalization from spinor representation of $O(3)$ to spin-1, spin-3/2 etc. 
Tensor  product of two generic representations $X_a$ and $X_b$  with different symmetries under $S_3$,
is a sum pf representation $X_{ab}$ each with symmetries being  $S, A, M^{\rho,\lambda}$. It is obvious that
\bea
&&S_a\otimes S_b=S_{ab}\nonumber\\
&&A_a\otimes A_b=S_{ab}\nonumber\\
&&S_a\otimes A_b=A_{ab}
\eea
but the product of the mixed representations $M^{\rho,\lambda}$ viewed as primitive $2_M$ doublets, is more subtle
and needs more detailed studies.

The product of two mixed representations $M^{\rho,\lambda}$ of $S_3$, yields the sum over the irreducible representations of $S_3$
\bea
\label{Y1}
\begin{ytableau}
~ & ~\\~\\
\end{ytableau}
\otimes
\begin{ytableau}
~&~\\~\\
\end{ytableau}
\sim
\begin{ytableau}
~\\
\end{ytableau}
\otimes
\begin{ytableau}
~\\
\end{ytableau}
=
\begin{ytableau}
~ \\~\\
\end{ytableau}
\oplus
\begin{ytableau}
~ & ~\\
\end{ytableau}
\nonumber\\
\eea
corresponding to the $1_A$ a singlet anti-symmetric, and to the $3_{S}$  triplet symmetric representations, respectively.

The singlet $1_A$ antisymmetric representation is 
\bea
A_{ab}=\frac 1{\sqrt 2} M_a^T
\begin{pmatrix}
0&1\\
-1 & 0
\end{pmatrix}
M_b=\frac 1{\sqrt 2}(M_a^\rho M_b^\lambda-M_a^\lambda M_b^\rho)\nonumber\\
\eea
while one of the triplet $3_S$  is e.g.
\bea
M_{ab}^\rho=\frac 1{\sqrt 2} M_a^T
\begin{pmatrix}
0&1\\
1 & 0
\end{pmatrix}
M_b=\frac 1{\sqrt 2}(M_a^\rho M_b^\lambda+M_a^\lambda M_b^\rho)\nonumber\\
\eea
The remaining part of the triplet 
$3_S$ with projection-$\pm$, are regrouped in the manifestly orthogonal combinations
\bea
S_{ab}=\frac 1{\sqrt 2} M_a^T
\begin{pmatrix}
1&0\\
0 & 1
\end{pmatrix}
M_b=\frac 1{\sqrt 2}(M_a^\rho M_b^\rho+M_a^\lambda M_b^\lambda)\nonumber\\
\eea
\bea
M_{ab}^\lambda=\frac 1{\sqrt 2} M_a^T
\begin{pmatrix}
1&0\\
0 & -1
\end{pmatrix}
M_b=\frac 1{\sqrt 2}(M_a^\rho M_b^\rho-M_a^\lambda M_b^\lambda)\nonumber\\
\eea
$S_{ab}$ is invariant under all the six rotations, hence all the six permutations. It  is   manifestly symmetric.
The  combinations $M^{\rho,\lambda}_{ab}$ can be checked to transform as a doublet under all
permutations, e.g.
\bea
{[P_4=(23)]}
\begin{pmatrix}
M_{ab}^\rho\\
M_{ab}^\lambda
\end{pmatrix}
&=&
\begin{pmatrix}
\frac 12&\frac{\sqrt 3}2\\
\frac{\sqrt{3}}2& -\frac 12
\end{pmatrix}
\begin{pmatrix}
M_{ab}^\rho\\
M_{ab}^\lambda
\end{pmatrix}\nonumber\\
\eea

  \begin{figure}[h]
        \centering
        \includegraphics[width=1.0\linewidth]{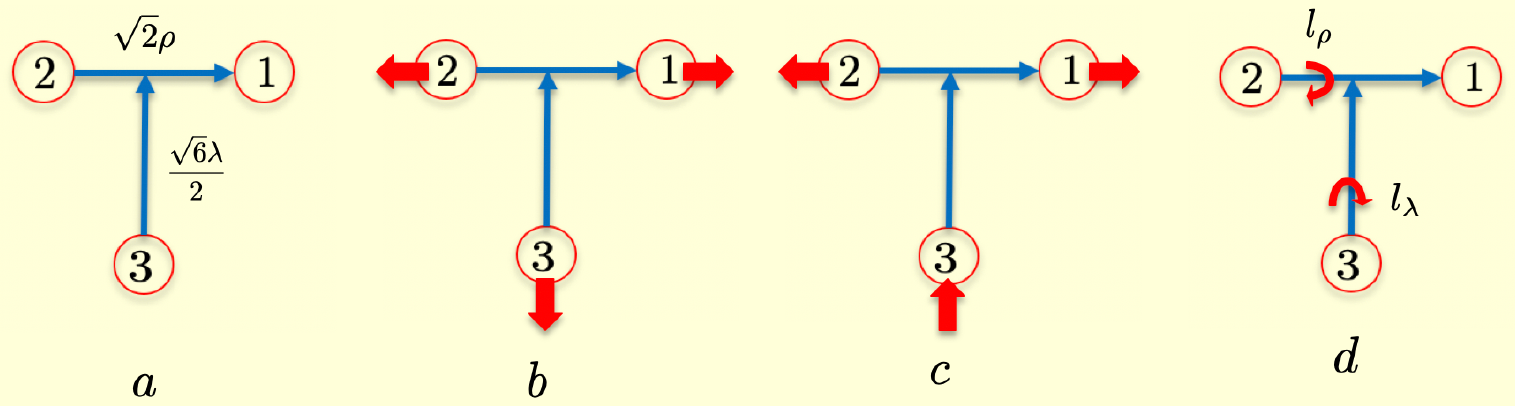}
        \caption{Three particle states for the nucleon in Jacobi coordinates: a) the ground state; b) the  radial monopole  excitation; c) the  radial tripole excitation; d) 
        the orbitally  excited states with internal $l_{\rho, \lambda}\neq 0$.}
        \label{fig_J123}
    \end{figure}

\subsection{Nucleon even parity states: radial excitations}
The lowest nucleon state with even-parity carries symmetric spin-flavor combination
 \bea
S_{ab}= \frac 1{\sqrt 2}(S_{a}^\rho F_{b}^\rho+S_{a}^\lambda F_{b}^\lambda)
 \eea
hence (spectroscopic notation $|LSJm\rangle_{X^P_R}$)
\bea \label{PWF0}
\bigg|0^+ \frac 12\frac 12 m\bigg\rangle_{p_S^+}= {\varphi_{1S}}(R)\frac 1{\sqrt{2}}(S_{ \frac 12 m}^\rho F_{\frac 12}^\rho+S_{\frac 12 m}^\lambda F_{\frac 12}^\lambda)
\eea
where the wave function $\varphi_{1S}(R)$ depends on the 6d hyperdistance $R=Y$,
with the squared norm
\bea
||0^+||^2=\int d\vec\rho\, d\vec\lambda\,|\varphi_{1S}(R)|^2
\eea

The radially excited states with zero total 
orbital angular momentum $L=l_\rho+l_\lambda=0$,  and null $l_\rho=l_\lambda=0$, but a total angular momentum $J=L+S=\frac 12$,
are illustrated in Fig.~\ref{fig_J123}b,c. These states are distinguishable  by their transformation under the permutation group. More
specifically, Fig.~\ref{fig_J123}b corresponds to the symmetric (S) and radially excited state 
\bea
\label{RADS}
\varphi_{2S}^S(R)=\frac 1{\sqrt 2} (R^2-\overline{R^2})\varphi_{1S}(R)
\eea
with a single node, that is orthogonal to ground state $\varphi_{1S}(R)$.

The states in Fig.~\ref{fig_J123}c are the radially excited scalar states with $\rho,\lambda$-mixed symmetry ($M^{\rho, \lambda}$),
\bea
\label{RADM}
\varphi_{3S}^\lambda(\rho,\lambda)&=&\frac 1{\sqrt{2}}(\rho\cdot\rho-\lambda\cdot \lambda)\varphi_{3S}(R)
\nonumber\\
\varphi_{3S}^\rho(\rho, \lambda)&=&\frac 1{\sqrt{2}}(\rho\cdot\lambda+\lambda\cdot \rho)\varphi_{3S}(R)
\eea

The lowest symmetric (S) and radially excited state of the  nucleon  follows from (\ref{RADS})
\bea 
\label{PWFRADS}
\bigg|0^{+*} \frac 12\frac 12 m\bigg\rangle_{p_S^{*+}}= \varphi_{2S}^S\frac 1{\sqrt{2}}(S_{ \frac 12 m}^\rho F_{\frac 12}^\rho+S_{\frac 12 m}^\lambda F_{\frac 12}^\lambda)\nonumber\\
\eea
which is orthogonal to $0^+$ with the squared norm
\bea
||0^{+*}||^2=\int d\vec\rho\,d\vec\lambda \,\frac 12(R^2-\overline{R^2})^2\,|\varphi_{1S}(R)|^2
\eea

The lowest mixed symmetry and radially excited state of the nucleon follows from (\ref{RADM})
\begin{widetext}
\bea 
\label{PWFRADM}
\bigg|0^+ \frac 12\frac 12 m\bigg\rangle_{p_M^{**+}}= \frac 1{{2}}
(F^\rho_{\frac 12}(S^\rho_{\frac 12 m}\varphi_{3S}^\lambda+S^\lambda_{\frac 12 m}\varphi_{3S}^\rho)+
F^\lambda_{\frac 12}(S^\rho_{\frac 12 m}\varphi_{3S}^\rho-S^\lambda_{\frac 12 m}\varphi^\lambda_{3S}))
\eea
with the first bracket carrying $M^\rho$ symmetry and the second bracket
carrying $M^\lambda$ symmetry. (\ref{PWFRADM}) is orthogonal to both $p^+,p^{+*}$ with squared norm
\bea
||0^{+**}||^2=\int d\vec\rho\,d\vec\lambda \,
\frac 14(\rho^4+\lambda^4-2\rho^2\lambda^2+4(\vec\lambda\cdot\vec\rho)^2)\,|\varphi_{3S}(R)|^2
\eea

\end{widetext}

\subsection{Nucleon even parity states: orbital excitations}
The orbitally excited states of the nucleon are $P,D$ states
with $L=1,2$ respectively, carry $l_{\rho,\lambda}\neq 0$
as illustrated in Fig.~\ref{fig_J123}d. For the S-state with $L=1$ or with $l_\rho=1$ and $l_\lambda=0$ and vice versa,
the spatial wavefunction with positive parity $1^+$,
is antisymmetric under the permutation group (A)
\bea
\label{L1}
\varphi^A_{11}(\rho,\lambda)&=&
\frac 1{\sqrt{2}}(\rho_+\sqrt{2}\lambda_z-\sqrt{2}\rho_z\lambda_+)\varphi_{1P}(R)\nonumber\\
\varphi^A_{10}(\rho,\lambda)&=&
\frac 1{\sqrt{2}}(\sqrt 2\rho_z\sqrt{2}\lambda_z-\sqrt{2}\rho_z\sqrt 2\lambda_z)\varphi_{1P}(R)=0\nonumber\\
\varphi^A_{1-1}(\rho,\lambda)&=&
\frac 1{\sqrt{2}}(\rho_-\sqrt{2}\lambda_z-\sqrt{2}\rho_z\lambda_)\varphi_{1P}(R)\nonumber\\
\eea
In contrast, for the D-state with $L=2$ or with $l_{\rho\lambda}=0,1,2$,
the spatial wavefunctions with positive parity $2^+$ and maximum weight,
split into symmetric (S) and mixed symmetry ($M^{\lambda, \rho}$) states
are tensor $\rho,\lambda$ combinations 
\begin{widetext}
\bea
\label{D2}
\varphi^S_{22}(\rho,\lambda)&=&
\frac 1{\sqrt{2}}(\rho_+^2+\lambda_+^2)\varphi_{1D}
=\sqrt{\frac{16\pi}{15}}(\rho^2Y_2^2(\hat\rho)+\lambda^2Y_2^2(\hat\lambda))\varphi_{1D}
\nonumber\\
\varphi^\lambda_{22}(\rho,\lambda)&=&
\frac 1{\sqrt{2}}(\rho_+^2-\lambda_+^2)\varphi_{1D}
=\sqrt{\frac{16\pi}{15}}(\rho^2Y_2^2(\hat\rho)-\lambda^2Y_2^2(\hat\lambda))\varphi_{1D}
\nonumber\\
\varphi^\rho_{22}(\rho,\lambda)&=&
\frac 1{\sqrt{2}}(\rho_+\lambda_++\lambda_+\rho_+)\varphi_{1D}
=\frac{4\pi}{3\sqrt 2}(Y_1^1(\hat\rho)Y_1^1(\hat\lambda)+
Y_1^1(\hat\lambda)Y_1^1(\hat\rho))
\rho\lambda\varphi_{1D}(R)
\eea

The P-excited states of the  nucleon  with $L^\pi=1^+$ and positive parity, follow from (\ref{L1})
\bea 
\label{PWFL=1}
\bigg|1^+ \frac 12\frac 12 m\bigg\rangle_{p_A^{+}}= C^{\frac 12 m}_{1m_L\frac 12 m_S}\varphi^A_{1m_L}\frac 1{\sqrt{2}}(S_{ \frac 12 m_S}^\lambda F_{\frac 12}^\rho-S_{\frac 12 m_S}^\rho F_{\frac 12}^\lambda)\nonumber\\
\eea
They are orthogonal to the S-states, with  the squared norm
\bea
||1^+||^2=\bigg\langle 1^+ \frac 12\frac 12 m\bigg|1^+ \frac 12\frac 12 m'\bigg\rangle=
\delta_{mm'}\,\int d\vec\rho\,d\vec\lambda\,\frac 13\rho^2\lambda^2\,|\varphi_{1P}(R)|^2
\eea
The D-excited states of the  nucleon  with $L^\pi=2^+$ and positive parity, follow from (\ref{D2})
\bea 
\label{PWFD=2}
\bigg|2^+ \frac 12\frac 12 m\bigg\rangle_{p_S^{+}}=
C^{\frac 12 m}_{2m_L\frac 32 m_S}S^S_{\frac 32m_S}
\frac 1{\sqrt{2}}( F_{\frac 12}^\rho\varphi^\rho_{2m_L}+
F_{\frac 12}^\lambda\varphi_{2m_L}^\lambda)\nonumber\\
\eea
They are orthogonal to the S,P-states, with  the squared norm
\bea
||2^+||^2=\bigg\langle 2^+ \frac 12\frac 12 m\bigg|2^+ \frac 12\frac 12 m'\bigg\rangle=
\delta_{mm'}\,\int d\vec\rho\,d\vec\lambda\,
\bigg(\frac 1{15}(\rho^4+\lambda^4)+\frac 29\,\rho^2\lambda^2\bigg)\,|\varphi_{1D}(R)|^2
\eea

To summarize: the positive parity nucleon state will be assumed to be
an admixture   of the 5 states
    \bea
    \label{FIVEX}
\bigg[\bigg|0^+ \frac 12\frac 12 m\bigg\rangle_{p_S^{+}}, 
\bigg|0^+ \frac 12\frac 12 m\bigg\rangle_{p_S^{*+}},
\bigg|0^+ \frac 12\frac 12 m\bigg\rangle_{p_M^{**+}}\bigg],
\bigg|1^+ \frac 12\frac 12 m\bigg\rangle_{p_A^{+}},
\bigg|2^+ \frac 12\frac 12 m\bigg\rangle_{p_S^{+}}    
    \eea
\end{widetext}
where the ground state $p^+_S$ is the dominant contribution.

\subsection{Spin-spin interaction}
The pair spin-spin interactions 
\bea
\label{SSX}
\mathbb V_{SS}(1,2,3)=\sum_{i<j=1}^3\,V_{SS}(i,j)\,
S_i\cdot S_j
\eea
split the degeneracy of the even-parity  states of the nucleon in (\ref{FIVEX}).  Since the states have good permutation symmetry, 
it is enough to evaluate the 12-pair contribution
\bea
V_{SS} (\sqrt{2} \rho)\,S_1\cdot S_2 \,
\eea
and multiply the net result by 3. 
We now detail some of the spin-spin matrix elements between the
states (\ref{FIVEX}), and summarize the final results.

In the nucleon ground state $p^+_S$ we have
\begin{widetext}
\bea
\bigg<0^+ \frac 12\frac 12 m\bigg|
\mathbb V_{SS}
\bigg|0^+ \frac 12\frac 12 m\bigg\rangle=&&
\frac 32 \bigg(S_{\frac 12 m}^\rho S_1\cdot S_2S_{\frac 12 m}^\rho+
S_{\frac 12 m}^\lambda S_1\cdot S_2S_{\frac 12 m}^\lambda\bigg)
\int d\vec\rho\,d\vec\lambda\,V_{SS}(\sqrt 2\rho)\,|\varphi_{1S}(R)|^2\nonumber\\
=&&
\frac 32\bigg(-\frac 34+\frac 14\bigg)\,\int d\vec\rho\,d\vec\lambda\,V_{SS}(\sqrt 2\rho)\,|\varphi_{1S}(R)|^2\equiv 
-\frac 34 \langle V_{1S1S}\rangle
\eea
\end{widetext}
with the overall factor of 3 counting the number of pairs.
Similarly in the nucleon D-state $p_S^+$, we note that
\bea
S_1\cdot S_2\bigg|2^+ \frac 12\frac 12 m\bigg\rangle_{p_S^+}=\frac 14
\bigg|2^+ \frac 12\frac 12 m\bigg\rangle_{p_S^+}
\eea
hence
\bea
\bigg<2^+ \frac 12\frac 12 m\bigg|
\mathbb V_{SS}
\bigg|2^+ \frac 12\frac 12 m\bigg\rangle=\frac 34 \langle V_{1S1S}\rangle
\eea
The final result for the spin-spin interaction in the positive parity nucleon basis-set (\ref{FIVEX}) is diagonal, and in agreement with (\ref{eqn_Hsd})
\begin{widetext}
\bea
\mathbb V_{SS}=-\frac 34\,
\begin{pmatrix}
\langle V_{1S1S}\rangle &0&0&0&0\\
0 &\langle V_{2S2S}\rangle&0&0&0\\
0 &0&\langle V_{3S3S}\rangle&0&0\\
0 &0&0&\langle V_{1L1L}\rangle&0\\
0 &0&0&0&-\langle V_{1D1D}\rangle\\
\end{pmatrix}
\eea
\end{widetext}

\subsection{Tensor interaction}
 Permutation symmetry reduces the tensor matrix elements
to the evaluation of  the 12-pair contribution
\bea
\label{TT12}
V_{T} (\sqrt{2} \rho)\,
(3S_1\cdot \hat{\rho}S_2\cdot \hat{\hat\rho} -S_1\cdot S_2)
\eea
with the net result multiplied by 3. In the spin
basis set 
\bea
[\alpha]=S^S_{\frac 32m_S=\pm \frac 32,\pm \frac 12}, S^\rho_{\frac  12 m_\rho=\pm \frac 12},  S^\lambda_{\frac 12 m_\lambda=\pm \frac 12}
\eea
the entries $[\alpha,\beta]$ of the $8\times 8$ spin-tensor part in (\ref{TT12}) are captured by the block-form matrix
\begin{widetext}
\bea
\mathbb V_{TS}[\alpha,\beta]=\sqrt{\frac \pi 5}\,
\begin{pmatrix}
& &+\frac 3{2_S} & +\frac 1{2_S}&+\frac 1{2_\lambda}&+\frac 1{2_\rho}
&-\frac 1{2_\rho}&-\frac 1{2_\lambda}&-\frac 1{2_S}&-\frac 3{2_S}\\
&&&&&&&&\\
+\frac 3{2_S} &&Y_2^0&\sqrt{2}Y_2^{-1}&Y_2^{-1}&0&0&2Y_2^{-2}&\sqrt{2}Y_2^{-2}&0\\
+\frac 1{2_S}&&-\sqrt{2}Y_2^1&-Y_2^0&-Y_2^0&0&0&-\sqrt{3}Y_2^{-1}&0&\sqrt{2}Y_2^{-2}\\
+\frac 1{2_\lambda}&&-Y_2^1&-\sqrt{2}Y_2^0&0&0&0&0&-\sqrt 3Y_2^{-1}&-2Y_2^{-2}\\
+\frac 1{2_\rho}&&0&0&0&0&0&0&0&0\\
-\frac 1{2_\rho}&&0&0&0&0&0&0&0&0\\
-\frac 1{2_\lambda}&&2Y_2^2&\sqrt 3Y_2^1&0&0&0&0&\sqrt 2Y_2^0&Y_2^{-1}\\
-\frac 1{2_S}&& \sqrt 2Y_2^2&0&\sqrt 3Y_2^1&0&0&\sqrt 2Y_2^0&-Y_2^0&-\sqrt 2Y_2^{-1}\\
-\frac 3{2_S}&&0&\sqrt 2Y_2^2&-2Y_2^2&0&0&-Y_2^1&\sqrt 2 Y_2^1&Y_2^0
\label{STT12}
\end{pmatrix}
\eea
\end{widetext}
with all spherical harmonics valued with $\hat\rho$. Throughout the calculations of the tensor transition matrix elements to follow and to avoid the multiplicatiion of notations, we will use the same labels  for the {\it normalized} radial wavefunctions, e.g. 
 $$\varphi_{1P}/||1^+||\rightarrow \varphi_{1P},\\\,\,\,\,\varphi_{1D}/||2^+|| \rightarrow \varphi_{1D}\,\,\,\,{\rm etc.}$$

\subsubsection{1S1D, 2S1D}
The tensor induced S-D mixing is given by the
transition matrix element
\begin{widetext}
\bea
\label{TSD1}
\bigg<0^+ \frac 12\frac 12 \frac 12\bigg|
\mathbb V_{T}
\bigg|2^+ \frac 12\frac 12 \frac 12\bigg\rangle=
3\,C^{\frac 12\frac 12}_{2m_L\frac 32 m_S}
\int d\vec\rho\,d\vec\lambda \,
\bigg(\varphi_{1S}\varphi_{2m_L}^\lambda  V_{T}(\sqrt 2\rho)\bigg)
\frac 12 S^\lambda_{\frac 12\frac 12}\mathbb V_{TS}^\rho S_{\frac 32m_S}
\eea
The $\lambda$-tensor wavefunction selected by the non-vanishing Clebsches, 
\bea
\label{CBS}
C^{\frac 12\frac 12}_{22\frac 32 -\frac 32}=+\sqrt{\frac 4{10}},\qquad
C^{\frac 12\frac 12}_{21\frac 32 -\frac 12}=-\sqrt{\frac 3{10}},\qquad
C^{\frac 12\frac 12}_{20\frac 32 \frac 12}=+\sqrt{\frac 2{10}},\qquad
C^{\frac 12\frac 12}_{2-1\frac 32 \frac 32}=-\sqrt{\frac 1{10}},
%
\eea
can be recast in spherical harmonics
\bea
\label{PHIX}
\varphi_{2-1}^\lambda&=&+\sqrt{\frac {8\pi}{15}}
(\rho^2Y_2^{-1}(\hat\rho)-\lambda^2Y_2^{-1}(\hat\lambda))\varphi_{1D}(R)\nonumber\\
\varphi_{20}^\lambda&=&+\sqrt{\frac {8\pi}{15}}
(\rho^2Y_2^{0}(\hat\rho)-\lambda^2Y_2^{0}(\hat\lambda))\varphi_{1D}(R)\nonumber\\
\varphi_{21}^\lambda&=&+\sqrt{\frac {8\pi}{15}}
(\rho^2Y_2^{1}(\hat\rho)-\lambda^2Y_2^{1}(\hat\lambda))\varphi_{1D}(R)\nonumber\\
\varphi_{22}^\lambda&=&+\sqrt{\frac {16\pi}{15}}
(\rho^2Y_2^{2}(\hat\rho)-\lambda^2Y_2^{2}(\hat\lambda))\varphi_{1D}(R)
\eea
Inserting (\ref{CBS}-\ref{PHIX}) into the transition matrix (\ref{TSD1}), using the pertinent spin-tensor entries in (\ref{STT12}), we obtain
\end{widetext}
\bea
&&\bigg<0^+ \frac 12\frac 12 \frac 12\bigg|
\mathbb V_T
\bigg|2^+ \frac 12\frac 12 \frac 12\bigg\rangle=
-\frac{3}{10\sqrt{15}}
\langle V_{1S1D}\rangle\nonumber\\
\eea
with
\bea
\langle V_{1S1D}\rangle=\int\,
d\vec\rho\,d\vec\lambda \,\rho^2\,\varphi_{1S}(R)\varphi_{1D}(R)\,V_T(\sqrt 2\rho)\nonumber\\
\eea
A similar result holds for the $2S1D$ mixing, with the replacement $\varphi_{1S}\rightarrow\varphi_{2S}$,
\bea
&&\bigg<0^{+*}\frac 12\frac 12 \frac 12\bigg|
\mathbb V_T
\bigg|2^+ \frac 12\frac 12 \frac 12\bigg\rangle=
-\frac{3}{10\sqrt{15}}
\langle V_{2S1D}\rangle\nonumber\\
\eea
with
\bea
\langle V_{2S1D}\rangle=\int\,
d\vec\rho\,d\vec\lambda \,\rho^2\,\varphi_{2S}(R)\varphi_{1D}(R)\,V_T(\sqrt 2\rho)\nonumber\\
\eea

\subsubsection{3S1D}
For the 3S-1D mixing, we have
\begin{widetext}
\bea
\label{TSD3}
&&\bigg<0^{+**} \frac 12\frac 12 m\bigg|
\mathbb V_{T}
\bigg|2^+ \frac 12\frac 12 m\bigg\rangle=3
C^{\frac 12 m}_{2m_L\frac 32 m_S}\int d\vec\rho\,d\vec \lambda\,\nonumber\\
&&\times\bigg(\frac 12(F^\rho_{\frac 12}(S^\rho_{\frac 12 m}\varphi^\lambda_{3S}+S^\lambda_{\frac 12 m}\varphi^\rho_{3S})
+F^\lambda_{\frac 12}(S^\rho_{\frac 12 m}\varphi^\rho_{3S}-S^\lambda_{\frac 12 m}\varphi^\lambda_{3S}))^*V_T^\rho
S_{\frac 32 m_S}^S\,
\frac 1{\sqrt{2}}(F_{\frac 12}^\rho\varphi^\rho_{2 m_L}+
F_{\frac 12}^\lambda\varphi^\lambda_{2 m_L})\bigg)
\eea
which reduces to
\bea
\label{3S1DX}
&&\bigg<0^{+**} \frac 12\frac 12 m\bigg|
\mathbb V_{T}
\bigg|2^+ \frac 12\frac 12 m\bigg\rangle=
\frac {3}{2\sqrt 2}C^{\frac 12 m}_{2m_L\frac 32 m_S}
\int d\vec\rho\,d\vec \lambda\,
(\varphi^{\rho*}_{3S}\varphi^\rho_{2m_L}
-\varphi^{\lambda*}_{3S}\varphi^{\lambda}_{2m_L})\,
S_{\frac 12 m}^\lambda V_T^\rho S_{\frac 32 m_S}^S
\eea
The $\rho,\lambda$-tensor wavefunctions selected by the non-vanishing Clebsches (\ref{CBS}), 
can be recast in spherical harmonics using (\ref{PHIX}) and
\bea
\label{PHIXX}
\varphi_{2-1}^\rho&=&+\frac{4\pi\sqrt 2}{3}
(Y_1^{-1}(\hat\rho)Y_1^0(\hat\lambda)+
Y_1^{-1}(\hat\lambda)Y_1^0(\hat\rho))\rho\lambda\varphi_{1D}(R)\nonumber\\
\varphi_{20}^\rho&=&+
\frac{4\pi\sqrt 2}3
(Y_1^0(\hat\rho)Y_1^0(\hat\lambda)+Y_1^0(\hat\lambda)Y_1^0(\hat \rho))\rho\lambda\varphi_{1D}(R))\nonumber\\
\varphi_{21}^\rho&=&+\frac{4\pi\sqrt 2}3
(Y_1^1(\hat\rho)Y_1^0(\hat\lambda)+Y_1^1(\hat\lambda)Y_1^0(\hat \rho))\rho\lambda\varphi_{1D}(R))\nonumber\\
\varphi_{22}^\rho&=&+
\frac{4\pi\sqrt 2}{3}
(Y_1^1(\hat\rho)Y_1^1(\hat\lambda)+Y_1^1(\hat\lambda)Y_1^1(\hat \rho))\rho\lambda\varphi_{1D}(R))
\eea
with the  3S-radial wavefunctions given in (\ref{RADM}), i.e.
\bea
\label{RADMX}
\varphi_{3S}^\lambda(\rho,\lambda)&=&\frac 1{\sqrt{2}}(\rho^2-\lambda^2)\varphi_{3S}(R)
\nonumber\\
\varphi_{3S}^\rho(\rho, \lambda)&=&\frac 1{\sqrt{2}}(\rho^i\lambda^i+\lambda^i\rho^i)\varphi_{3S}(R)=
\frac{4\pi\sqrt 2}{3}(Y_{1}^{1*}(\hat\rho)Y_1^1(\hat\lambda)+Y_1^{1*}(\hat\lambda)Y_1^1(\hat\rho)+Y_1^0(\hat\rho)Y_1^0(\hat\lambda))\rho\lambda\varphi_{3S}(R)\nonumber\\
\eea
The spin-tensor matrix elements in (\ref{3S1DX}) can be read off (\ref{STT12}),
with the result
\bea
\label{ST12X}
S_{\frac 12 \frac 12}^\lambda V_T^\rho S_{\frac 32 m_S}^S=-\delta_{m_S+m_L,\frac 12}\sqrt{\frac{(2+m_L)\pi}{5}}Y_2^{-m_L}
\eea
for the selected  $m_L=2,1,0,-1$ and fixed $m=\frac 12$.
Inserting (\ref{CBS}-\ref{ST12X}) into the transition matrix (\ref{3S1DX}), we obtain 
\bea
\label{3S1DXFIN}
\bigg<0^{+**} \frac 12\frac 12 m\bigg|
\mathbb V_{T}\bigg|2^+ \frac 12\frac 12 m\bigg\rangle=&&
-\frac 1{5\sqrt{15}}{(2+\sqrt 2)}
\int d\vec\rho \,d\vec\lambda\, \rho^2\,\lambda^2\,\varphi_{3S}(R)\varphi_{1D}(R)\nonumber\\
&&+\frac 1{6\sqrt{15}}
\int d\vec\rho \,d\vec\lambda\, (\rho^2-\lambda^2)\rho^2\,\varphi_{3S}(R)\varphi_{1D}(R)
\eea
with the first contribution stemming from the $\rho\rho$ term, and the second contribution from
the $\lambda\lambda$ term.

\subsubsection{1P1D}
A similar analysis yields the tensor P-D  matrix element
\bea
\label{PDXX}
\bigg<1^+ \frac 12\frac 12 \frac 12\bigg|
\mathbb V_T
\bigg|2^+ \frac 12\frac 12 \frac 12\bigg\rangle=&&\frac 32 C^{\frac 12\frac 12}_{1m_L\frac 12 m_S}
C^{\frac 12\frac 12}_{2\bar m_L\frac 32 \bar m_S}\nonumber\\
&&\times\int\,d\vec\rho\,d\vec \lambda\,
(\varphi^{A*}_{1m_L}\varphi^\rho_{2\bar m_L}\,S^\lambda_{\frac 12m_S}V_T^\rho S_{\frac 32\bar m_S}^S
-\varphi^{A*}_{1m_L}\varphi^\lambda_{2\bar m_L}\,S^\rho_{\frac 12m_S}V_T^\rho S_{\frac 32\bar m_S}^S)\nonumber\\
\eea
From (\ref{STT12}) it follows that  all entries in $S^\rho_{\frac 12m_S}V_T^\rho S_{\frac 32\bar m_S}^S=0$, and (\ref{PDXX}) simplifies
\bea
\label{PDYY}
\bigg<1^+ \frac 12\frac 12 \frac 12\bigg|
\mathbb V_T
\bigg|2^+ \frac 12\frac 12 \frac 12\bigg\rangle=\frac 32 C^{\frac 12\frac 12}_{1m_L\frac 12 m_S}
C^{\frac 12\frac 12}_{2\bar m_L\frac 32 \bar m_S}
\int\,d\vec\rho\,d\vec \lambda\,
\varphi^{A*}_{1m_L}\varphi^\rho_{2\bar m_L}\,
S^\lambda_{\frac 12m_S}V_T^\rho S_{\frac 32\bar m_S}^S
\eea
The $1^+$-antisymmetric wavefunctions in (\ref{L1}) can be recast in spherical harmonics
\bea
\label{EXPXX}
\varphi_{11}^A&=&\sqrt{\frac {4\pi}3}
(Y_1^1(\hat\rho)Y_1^0(\hat\lambda)-Y_1^0(\hat\rho)Y_1^1(\hat\lambda))\,\rho\lambda\,\varphi_{1P}(R)\nonumber\\
\varphi_{10}^A&=&\sqrt{\frac {4\pi}3}
(Y_1^0(\hat\rho)Y_1^0(\hat\lambda)-Y_1^0(\hat\rho)Y_1^0(\hat\lambda))\,\rho\lambda\,\varphi_{1P}(R)=0\nonumber\\
\varphi_{1-1}^A&=&\sqrt{\frac {4\pi}3}
(Y_1^{-1}(\hat\rho)Y_1^0(\hat\lambda)-Y_1^0(\hat\rho)Y_1^{-1}(\hat\lambda))\,\rho\lambda\,\varphi_{1P}(R)
\eea
The reduction of (\ref{PDYY}) using the explicit wavefunctions (\ref{PHIXX}, \ref{EXPXX}) is tedious and will not be detailed here.

\subsubsection{1D1D}
Finally, the  tensor D-D matrix element
\bea
\label{TDD1}
\bigg<2^+ \frac 12\frac 12 m\bigg|
\mathbb V_{T}
\bigg|2^+ \frac 12\frac 12 m\bigg\rangle=&&3C^{\frac 12 m}_{2m_L\frac 32 m_S}
C^{\frac 12 m}_{2\bar m_L\frac 32 \bar m_S}\nonumber\\
&&\times\int d\vec\rho\,d\vec\lambda \,
\frac 1{\sqrt{2}}(F_{\frac 12}^\rho\varphi^\rho_{2m_L}+
F_{\frac 12}^\lambda\varphi^\lambda_{2m_L})^*\,
S^S_{\frac 32m_S}V_T^\rho S_{\frac 32\bar m_S}^S\,
\frac 1{\sqrt{2}}(F_{\frac 12}^\rho\varphi^\rho_{2\bar m_L}+
F_{\frac 12}^\lambda\varphi^\lambda_{2\bar m_L})\nonumber\\
=&&\frac 32 C^{\frac 12 m}_{2m_L\frac 32 m_S}
C^{\frac 12 m}_{2\bar m_L\frac 32 \bar m_S}
\int d\vec\rho\,d\vec\lambda \,
(\varphi^{\rho *}_{2m_L}\varphi^\rho_{2\bar m_L}
+\varphi^{\lambda *}_{2m_L}\varphi^\lambda_{2\bar m_L})\,
S^S_{\frac 32m_S}V_T^\rho S_{\frac 32 \bar m_S}^S\,
\eea
simplifies by  angular integration. The reduction of (\ref{TDD1}) is also tedious, and for illustration of the procedure we present only the contribution stemming from the diagonal spin-tensor entries in (\ref{TDD1}). Using the spin-tensor entries in (\ref{STT12}) we have  
\bea
S^S_{\frac 32m_S}V_T^\rho S_{\frac 32 m_S}^S=
(-1)^{\frac 52+|m_S|}\sqrt{\frac \pi 5}\,Y_2^0(\hat \rho)
\eea
which for $m=\frac 12$ gives
\bea
\label{TDD2}
\frac 32 \sqrt{\frac \pi 5}\bigg(
|C^{\frac 12 \frac 12}_{22\frac 32 -\frac 32}|^2
\int d\vec\rho\,d\vec\lambda \,
|\varphi^\lambda_{22}|^2\,
\,Y_2^0(\hat \rho)-
|C^{\frac 12 \frac 12}_{21\frac 32 -\frac 12}|^2
\int d\vec\rho\,d\vec\lambda \,
|\varphi^\lambda_{21}|^2\,
\,Y_2^0(\hat \rho)\bigg)\nonumber\\
\eea
Since the  mixed symmetry wavefunctions (\ref{D2}) can be recast in spherical harmonics
\bea
\varphi_{22}^\lambda=
\sqrt{\frac{16\pi}{15}}(\rho^2Y_2^2(\hat\rho)-\lambda^2Y_2^2(\hat\lambda))\varphi_{1D}
\qquad
\varphi_{21}^\lambda=
\sqrt{\frac{8\pi}{15}}(\rho^2Y_2^1(\hat\rho)-\lambda^2Y_2^1(\hat\lambda))\varphi_{1D}
\eea
and using the identities
\bea
\int d\hat\rho\,Y_2^{2*}(\hat \rho)Y_2^0(\hat \rho)\,Y_2^{2}(\hat \rho)=
2\int d\hat\rho\,Y_2^{1*}(\hat \rho)Y_2^0(\hat \rho)\,Y_2^{1}(\hat \rho)=
\frac 27 \sqrt{\frac 5{4\pi}}
\eea
the spin-diagonal contribution in (\ref{TDD1}) is
\bea
\label{TDD3}
\bigg(
|C^{\frac 12 \frac 12}_{22\frac 32 -\frac 32}|^2
-\frac 14
|C^{\frac 12 \frac 12}_{21\frac 32 -\frac 12}|^2\bigg)
\frac 2{35}\langle V_{1D1D}\rangle
=\bigg(\sqrt{\frac 15}^2-\frac 14\sqrt{\frac 3{10}}^2\bigg)\frac {2}{35}\langle V_{1D1D}\rangle
=\frac {1}{140}\langle V_{1D1D}\rangle
\eea
\end{widetext}
with
\bea
\langle V_{1D1D}\rangle=\int\,
d\vec\rho\,d\vec\lambda \,\rho^4\,\varphi_{1D}\varphi_{1D}\,V_T(\sqrt 2\rho)
\eea
The additional off-diagonal spin contributions  in (\ref{TDD1}), follow from the same reasoning and will not be presented.


\subsection{Details of $V_{SS}, V_T$ matrix elements}
The perturbative (one gluon exchange) expressions for the spin-dependent potentials are
\ba
\label{SPSP}
V_{SS}^{12}&=&{2 \alpha_s\over 3 m^2}{8\pi  \over 3}\delta^3\big(\sqrt{2}\vec \rho \big)  \\
V_{T}^{12}&=&{2 \alpha_s\over 3 m^2}{1 \over (\sqrt{2} \rho)^3} \nonumber
\ea
where the inter-quark distance  is $r_{12}=(\sqrt{2} \rho)$.
The matrix elements of the spin and tensor operators (\ref{SPSP}), over the wave functions include integration over all 6  coordinates $\vec \rho,\vec \lambda$. 

For the perturbative spin-spin term, we use the angular measure $d^3\rho \lambda^2 d\lambda \rm sin(\theta_\lambda) d\theta_\lambda \phi_\lambda$, with the result 
\ba 
&&\langle A | V_{SS} | B \rangle =3 {2 \alpha_s\over 3 m^2}{8\pi  \over 3}{1 \over  2^{3/2}} \nonumber \\
&&\times\int \psi^*_A(\lambda) (\vec S^1\cdot \vec S^2)\psi_B(\lambda)\, \lambda^2 d\lambda \rm sin(\theta_\lambda) d\theta_\lambda \phi_\lambda \nonumber\\
\ea

\begin{figure}[t!]
    \centering
    \includegraphics[width=0.95\linewidth]{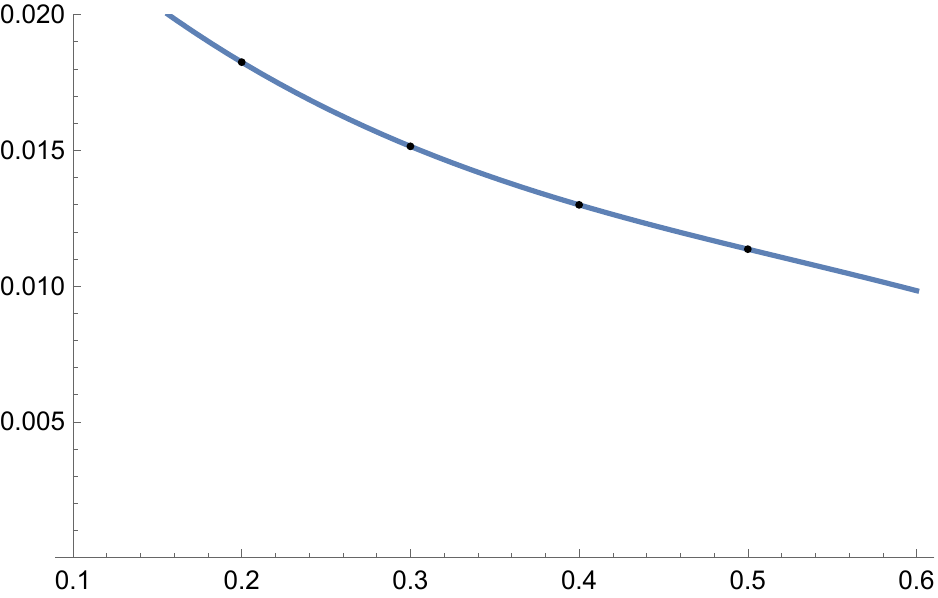}
    \caption{The dependence of the integral (\ref{eqn_T_of_delta})
    on the regulator $\delta$ (in $\rm GeV^{-1}$)}
    \label{fig_tensor_1s1d}
\end{figure}

The perturbative tensor force  needs to be regulated
$$ 1/r^3 \rightarrow 1/(r^2+\delta^2)^{3/2}$$
The regulator  $\delta$ is fixed by physical considerations.
First, for $\delta < 1/m_q$ we loose  the range of validity of the non-relativistic approximation. The resulting "zitterbewegung" can in principle be
remedied by the Dirac equation. Second, as we discussed in the section on form factors, the constituent quark is not a point-like object. Its form-factor
well agrees with that from the ILM,  which relates $\delta$ to the mean instanton size $\rho_{inst}\approx 0.3 \,\rm fm\sim 1.4\, \rm GeV^{-1}$. 
While in principle there remains some dependence on the regulator size,
this dependence is in practice quite weak. In Fig.~\ref{fig_tensor_1s1d}
we show  the transion matrix element of the perturbative tensor force between 1S and 1D state
\be \label{eqn_T_of_delta}
T(\delta)=\int {
\psi_{1S}^*(Y)\psi_{1D}(Y) \over (\rho^2+\lambda^2)^5 (\rho^2+\delta^2)^{3/2}
} {\rho^2 d\rho \lambda^2 d\lambda}
\ee
with $Y=\sqrt{\rho^2+\lambda^2}$.
The tensor operator producing this mixing has
a fixed structure, but its magnitude in the existing theory papers,
has relied mostly on the perturbative derivation.
Lacking a full non-perturbative derivation from first principles, we can
rely on various observables to quantify its magnitude.


\begin{widetext}
\section{Explicit orbit-spin-flavor wave functions in the ``monom representation"}
\label{eqn-orbit-spin-flavor}
Let us start 
 with the standard spin-isospin wave function  for the nucleon 
\be | 1S > =\psi_{1S}(Y) \big(S_\rho I_\rho+ 
S_\lambda I_\lambda \big) /\sqrt{2} 
 \ee 
in terms of   the spin and isospin operators defined in \ref{eqn_Srho}.
The 2S state is the same, with the appropriate radial wave function.
In the main part of the paper, we have used the decomposition into spin-flavor or 64 "monoms", writing
them as Tables in Mathematica with indices defined in the order $$\rm Sum[...\{s1,1,2\},\{s2,1,2\}\{s3,1,2\}\{f1,1,2\}\{f2,1,2\}\{f3,1,2\}] $$
the nucleon 1S and 2S states have the same angle-independent wave functions shown in Fig.\ref{fig_proton_nb}

\begin{figure}[h!]
    \centering
    \includegraphics[width=0.75\linewidth]{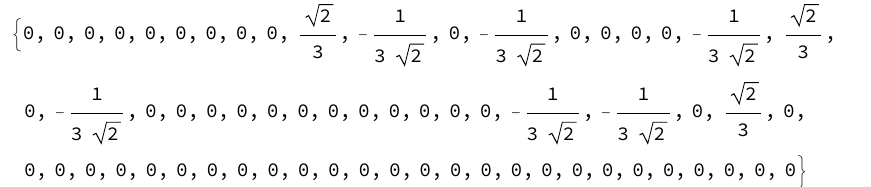}
    \caption{1S (proton) and 2S wave function in spin-flavor tensor form}
    \label{fig_proton_nb}
\end{figure}

Here only 9 out of 64 possible ``monom" terms are nonzero, so 
it may appear unnecessary to write it in this way.
However, with the orbital parts containing two powers of coordinates, this reporting is  very useful, as we now detail. Indeed, the construction of the
actual wave-functions can start with the  permutation
invariant combinations $V_{ABC},V1,V2,V3$ as defined
in Appendix \ref{sec_app_permuts}. This proceeds
through two steps: (i)  we combine 
a pair of standard $Y_{1,m}(\theta,\phi)$ functions into three combinations with $L=0,1,2$ (scalar, vector and tensor) using Clebsch-Gordon coefficients; (ii) we combine them  with spin operators (using another set of Clebsch-Gordon coefficients) to obtain states with 
total angular momentum $J=1/2$.
  

More specifically, let us start with the state with the largest angular momentum $L=2$. The $J=1/2$ can only be formed if the total spin is $S=3/2$, which are permutation symmetric operators
 \ba 
\rm spin[3/2,3/2]&=&\uparrow  \uparrow\uparrow \nonumber \\
\rm spin[3/2,1/2]&=&(\downarrow  \uparrow\uparrow + \uparrow\downarrow \uparrow +\uparrow\uparrow\downarrow)/\sqrt{3} \ea
and $\rm spin[3/2,-1/2],spin[3/2,3/2]$ defined by a reflection.
Using the Clebsch-Gordon coefficients, we now define the wave-function of the $L=2,S=3/2,J=1/2$``tensor" state as
 \ba \rm "tensor"=\frac 1{\sqrt{10}}\,
\bigg(&-&
   \rm  (-2 spin[3/2, -(3/2)] \psi[2,-2,\lambda\lambda] + 
     \sqrt{3} spin[3/2, -(1/2)] \psi[2,-1,\lambda\lambda] \nonumber \\
     &-& 
    \rm  \sqrt{2} spin[3/2, +(1/2)] \psi[2,0,\lambda\lambda] + 
     spin[3/2, +(3/2)] \psi[2,1,\lambda\lambda]*I_\lambda 
     \nonumber \\ &+&
 \rm  (-2 spin[3/2, 3/-2] \psi[2,-2,\rho\rho] + 
     \sqrt{3} spin[3/2, -(1/2)] \psi[2,-1,\rho\rho] \nonumber \\
     &-& 
    \rm \sqrt{2} spin[3/2, +(1/2)] \psi[2,0,\rho\rho] + 
     spin[3/2, +(3/2)] \psi[2,1,\rho\rho]*I_\lambda \nonumber \\ &+&
    \rm (-2 spin[3/2, -(3/2)] \psi[2,-2,\rho\lambda] + 
     \sqrt{3} spin[3/2, -1/2] \psi[2,-1,\rho\lambda]  \nonumber \\ &-& 
    \rm  \sqrt{2} spin[3/2, +(1/2)] \psi[2,0,\rho\lambda] + 
     spin[3/2, +(3/2)] \psi[2,1\rho\lambda]*I_\rho \nonumber \\ &+&
   \rm (-2 spin[3/2, -(3/2)] \psi[2,-2,\lambda\rho] + 
     \sqrt{3} spin[3/2, -(1/2)] \psi[2,-1,\lambda\rho] \nonumber \\  &-& 
    \rm  \sqrt{2} spin[3/2, +(1/2)] \psi[2,0,\lambda\rho] + 
     spin[3/2, +(3/2)] \psi[2,1,\lambda\rho]*I_\rho)\bigg)
 \ea
Here the coordinate part of the wave functions are two spin-1
harmonics $Y_{1m}(\theta_i,\phi_i)$, combined into $L=2$,
with the remaining pair of indices indicating their angular part. For example $\psi[2,-2,\lambda\lambda]$ means $L=2,L_z=-2$
combination and both angles are $\theta_\lambda,\phi_\lambda$.
In this explicit form $\lambda,\rho$ mean the length of the
corresponding vectors, with their angles shown explicitly. (Of course,
$\lambda,\rho$ can be expressed via the 5-th angle $\chi$, but we have not done so here for
brevity of notations.)
 Their (flatten) spin-tensor form is given in Figs \ref{fig_VL2_nb},
\ref{fig_VL2_nb_2}
below.

\begin{figure}[h!]
    \centering
\includegraphics[width=0.75\linewidth]{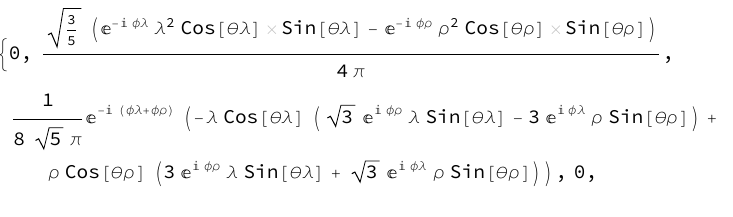}
\includegraphics[width=0.75\linewidth]{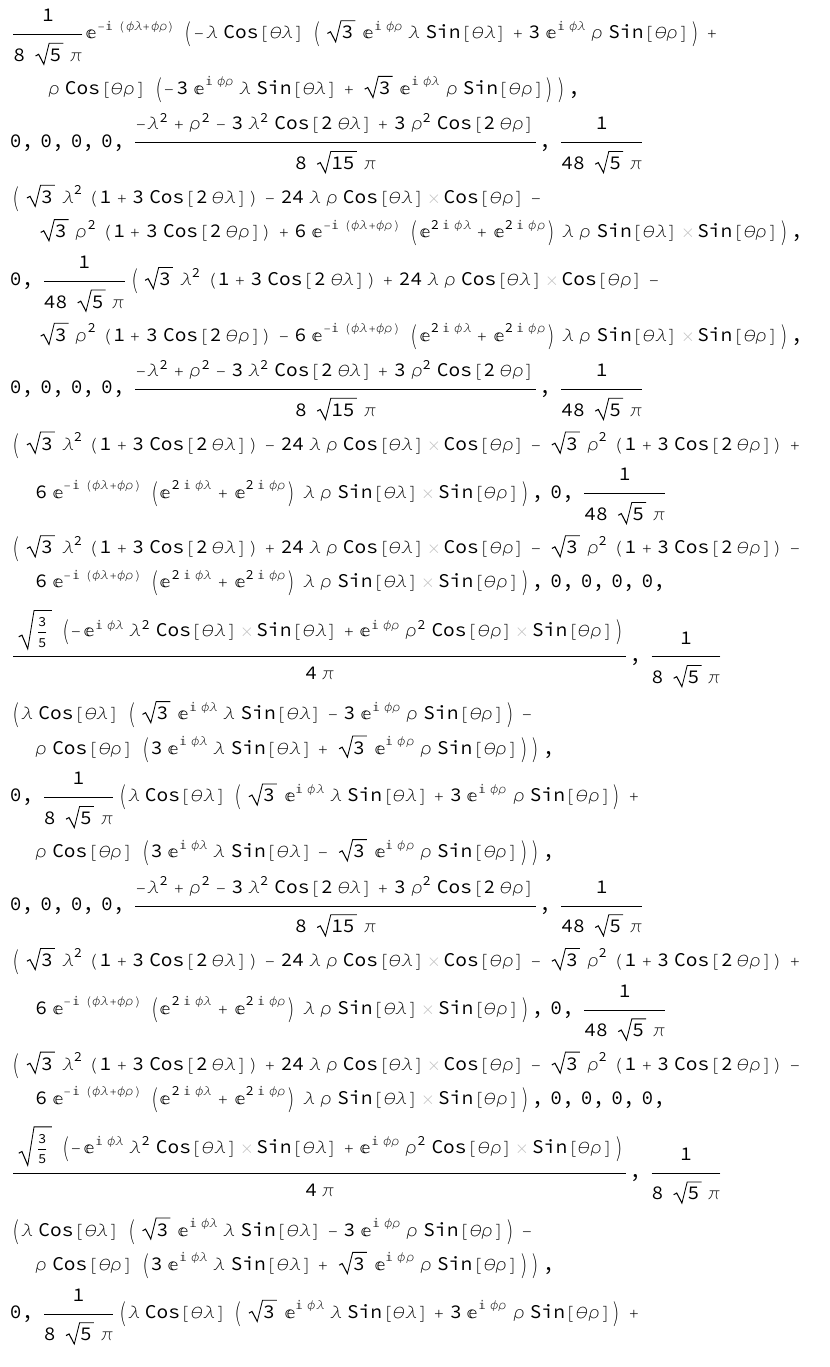}
    \label{fig_VL2_nb}
\end{figure}
\begin{figure}[h!]
    \centering
\includegraphics[width=0.75\linewidth]{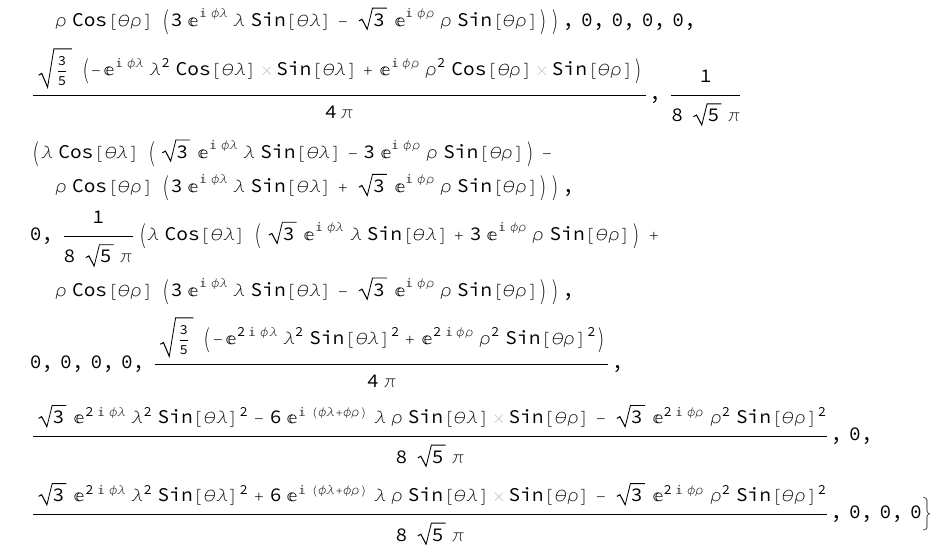}
\caption{Orbital-spin-flavor wave function $VL2$, or the ``tensor" state}
    \label{fig_VL2_nb_2}\end{figure}

The other cases start with the invariants $V1,V2,V3$
made of four $\rho,\lambda$ type objects. 
Not all of them may contribute. 
In particular, since  $A_\alpha B_\beta$
are just coordinates $\vec\rho,\vec\lambda$, it is convenient to start with
 $L=1$ corresponding  to the antisymmetric combination $[\vec \rho \times \vec \lambda] $ so that only $V3$  contributes.
The positive parity $L=1 $ state is shown in Fig.\ref{fig_VL1_nb}

 The $L=1,S=1/2,J=1/2$ ``vector" state is 
 \ba "\rm vector" &=&   (\rm (\psi[1,-1,\rho\lambda] - \psi[1,-1,\lambda\rho])  *(spin[\rho,-1/2] *
      I_\lambda - spin[\lambda,-1/2]*I_\rho) *Sqrt[2/3]  \nonumber \\
 & -& \rm (\psi[1,0,\rho\lambda] - \psi[1,0,\lambda\rho])  *(spin[\rho,1/2]*
      I_\lambda - spin[\lambda,1/2]*I_\rho)*1/Sqrt[3]) 
 \ea 
with  the $\rm spin[\rho,\pm 1/2]$ or $\rm spin[\lambda,\pm 1/2]$ are spin 1/2
operators of rho or lambda kind. $I_\rho$ and $I_\lambda$
are similar isospin operators, all with up (+1/2) components as
we discuss proton-like states.

\begin{figure}[h!]
    \centering
    \includegraphics[width=0.75\linewidth]{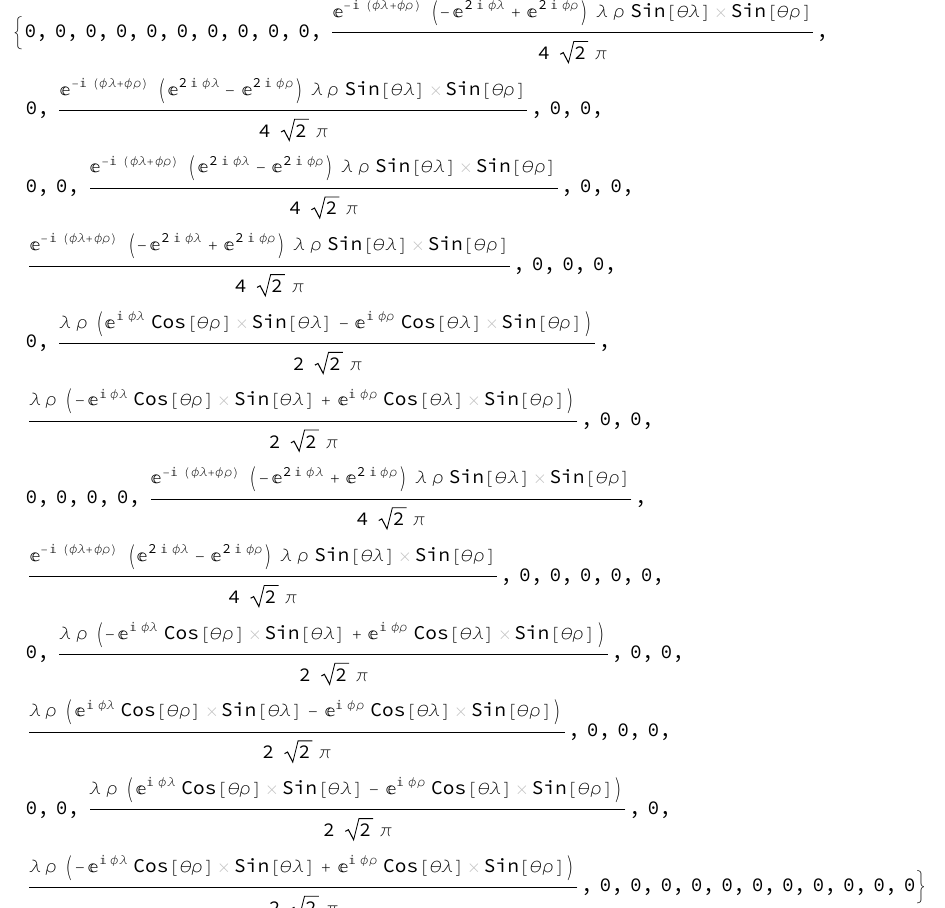}
    \caption{Orbital-spin-flavor wave function $VL1$, }
    \label{fig_VL1_nb}
\end{figure}

The last state $VL0$ is made out of two ``scalar" combinations of  $L=0$ possible states out of $V1,V2$. After they are formed, it turned out that they are not orthogonal to each other, and  after
ortho-gonalization we found that one combination 
has no angles, and is just another copy of the 2S wave
function. The physical one takes the form shown in Fig.\ref{fig_VL0_nb}.
\begin{figure}[h!]
    \centering
    \includegraphics[width=0.75\linewidth]{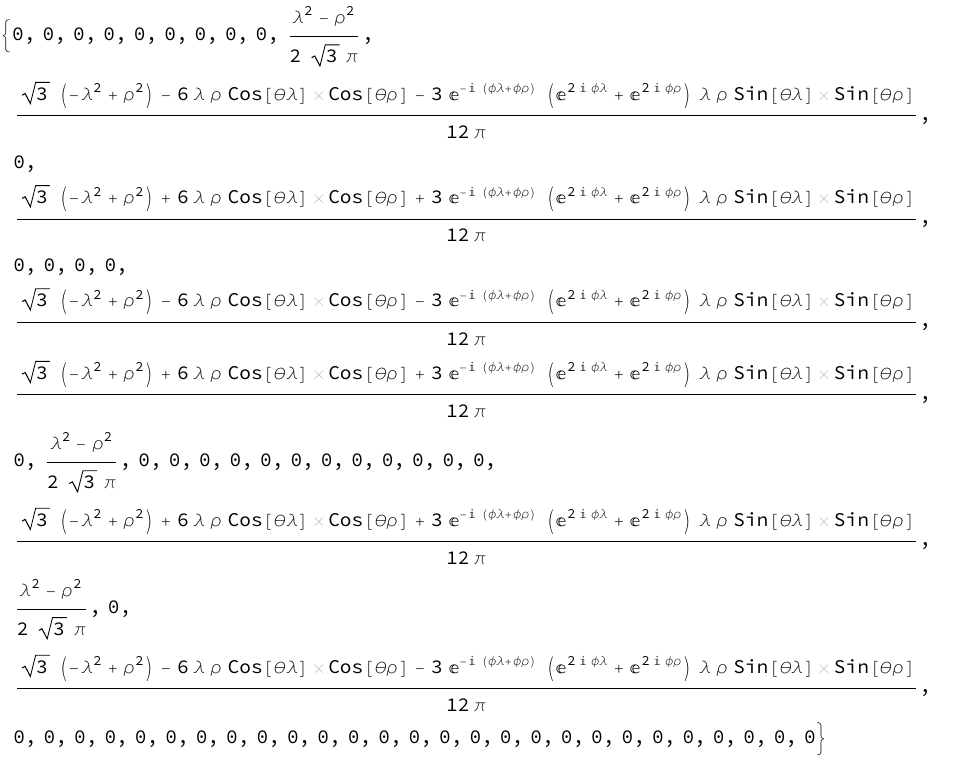}
    \caption{Orbital-spin-flavor wave function $VL0$,  or the ``scalar" state}
    \label{fig_VL0_nb}
\end{figure}
While these functions are not yet normalized,
they are orthogonal to each other. Multiple tests were
performed, among them with the operators $S_z$ and $L_z$,
the third components of the spin and orbital momentum.
Using 5 states in the order we perform averages with them, $| J >=\{1S,2S,VL0,VL1,VL2\}$, we have
the following
\be \langle S_z\rangle=\{ 1/2, 1/2, 1/2, -(1/6), -(1/2) \} \ee
\be \langle L_z\rangle=\{ 0, 0, 0, 2/3, 1 \} \ee
which adds correctly to $\langle J_z\rangle=1/2 $ for each of them.
The same test carries to the isospin operators  $I_z=1/2$, so the results  describe correctly the proton, and its excitations with the same global quantum number $J^P=(1/2)^+$.

\end{widetext}

\bibliography{orbital}

\begin{thebibliography}{25}%
\makeatletter
\providecommand \@ifxundefined [1]{%
 \@ifx{#1\undefined}
}%
\providecommand \@ifnum [1]{%
 \ifnum #1\expandafter \@firstoftwo
 \else \expandafter \@secondoftwo
 \fi
}%
\providecommand \@ifx [1]{%
 \ifx #1\expandafter \@firstoftwo
 \else \expandafter \@secondoftwo
 \fi
}%
\providecommand \natexlab [1]{#1}%
\providecommand \enquote  [1]{``#1''}%
\providecommand \bibnamefont  [1]{#1}%
\providecommand \bibfnamefont [1]{#1}%
\providecommand \citenamefont [1]{#1}%
\providecommand \href@noop [0]{\@secondoftwo}%
\providecommand \href [0]{\begingroup \@sanitize@url \@href}%
\providecommand \@href[1]{\@@startlink{#1}\@@href}%
\providecommand \@@href[1]{\endgroup#1\@@endlink}%
\providecommand \@sanitize@url [0]{\catcode `\\12\catcode `\$12\catcode
  `\&12\catcode `\#12\catcode `\^12\catcode `\_12\catcode `\%12\relax}%
\providecommand \@@startlink[1]{}%
\providecommand \@@endlink[0]{}%
\providecommand \url  [0]{\begingroup\@sanitize@url \@url }%
\providecommand \@url [1]{\endgroup\@href {#1}{\urlprefix }}%
\providecommand \urlprefix  [0]{URL }%
\providecommand \Eprint [0]{\href }%
\providecommand \doibase [0]{http://dx.doi.org/}%
\providecommand \selectlanguage [0]{\@gobble}%
\providecommand \bibinfo  [0]{\@secondoftwo}%
\providecommand \bibfield  [0]{\@secondoftwo}%
\providecommand \translation [1]{[#1]}%
\providecommand \BibitemOpen [0]{}%
\providecommand \bibitemStop [0]{}%
\providecommand \bibitemNoStop [0]{.\EOS\space}%
\providecommand \EOS [0]{\spacefactor3000\relax}%
\providecommand \BibitemShut  [1]{\csname bibitem#1\endcsname}%
\let\auto@bib@innerbib\@empty
\bibitem [{\citenamefont {Miesch}\ \emph {et~al.}(2023)\citenamefont {Miesch},
  \citenamefont {Shuryak},\ and\ \citenamefont {Zahed}}]{Miesch:2023hvl}%
  \BibitemOpen
  \bibfield  {author} {\bibinfo {author} {\bibfnamefont {Nicholas}\
  \bibnamefont {Miesch}}, \bibinfo {author} {\bibfnamefont {Edward}\
  \bibnamefont {Shuryak}}, \ and\ \bibinfo {author} {\bibfnamefont {Ismail}\
  \bibnamefont {Zahed}},\ }\bibfield  {title} {\enquote {\bibinfo {title}
  {{Hadronic structure on the light-front. IX. Orbital-spin-isospin wave
  functions of baryons}},}\ }\href {\doibase 10.1103/PhysRevD.108.094033}
  {\bibfield  {journal} {\bibinfo  {journal} {Phys. Rev. D}\ }\textbf {\bibinfo
  {volume} {108}},\ \bibinfo {pages} {094033} (\bibinfo {year} {2023})},\
  \Eprint {http://arxiv.org/abs/2308.14694} {arXiv:2308.14694 [hep-ph]}
  \BibitemShut {NoStop}%
\bibitem [{\citenamefont {Isgur}\ and\ \citenamefont
  {Karl}(1979)}]{Isgur:1978wd}%
  \BibitemOpen
  \bibfield  {author} {\bibinfo {author} {\bibfnamefont {Nathan}\ \bibnamefont
  {Isgur}}\ and\ \bibinfo {author} {\bibfnamefont {Gabriel}\ \bibnamefont
  {Karl}},\ }\bibfield  {title} {\enquote {\bibinfo {title} {{Positive Parity
  Excited Baryons in a Quark Model with Hyperfine Interactions}},}\ }\href
  {\doibase 10.1103/PhysRevD.19.2653} {\bibfield  {journal} {\bibinfo
  {journal} {Phys. Rev. D}\ }\textbf {\bibinfo {volume} {19}},\ \bibinfo
  {pages} {2653} (\bibinfo {year} {1979})},\ \bibinfo {note} {[Erratum:
  Phys.Rev.D 23, 817 (1981)]}\BibitemShut {NoStop}%
\bibitem [{\citenamefont {Miesch}\ and\ \citenamefont
  {Shuryak}(2024)}]{Miesch:2024vjk}%
  \BibitemOpen
  \bibfield  {author} {\bibinfo {author} {\bibfnamefont {Nicholas}\
  \bibnamefont {Miesch}}\ and\ \bibinfo {author} {\bibfnamefont {Edward}\
  \bibnamefont {Shuryak}},\ }\bibfield  {title} {\enquote {\bibinfo {title}
  {{Wave functions of multiquark hadrons from representations of the symmetry
  groups $S_n$}},}\ }\href@noop {} {\  (\bibinfo {year} {2024})},\ \Eprint
  {http://arxiv.org/abs/2406.05024} {arXiv:2406.05024 [hep-ph]} \BibitemShut
  {NoStop}%
\bibitem [{\citenamefont {Isgur}\ and\ \citenamefont
  {Karl}(1978)}]{Isgur:1978xj}%
  \BibitemOpen
  \bibfield  {author} {\bibinfo {author} {\bibfnamefont {Nathan}\ \bibnamefont
  {Isgur}}\ and\ \bibinfo {author} {\bibfnamefont {Gabriel}\ \bibnamefont
  {Karl}},\ }\bibfield  {title} {\enquote {\bibinfo {title} {{P Wave Baryons in
  the Quark Model}},}\ }\href {\doibase 10.1103/PhysRevD.18.4187} {\bibfield
  {journal} {\bibinfo  {journal} {Phys. Rev. D}\ }\textbf {\bibinfo {volume}
  {18}},\ \bibinfo {pages} {4187} (\bibinfo {year} {1978})}\BibitemShut
  {NoStop}%
\bibitem [{\citenamefont {Capstick}\ and\ \citenamefont
  {Roberts}(2000)}]{Capstick:2000qj}%
  \BibitemOpen
  \bibfield  {author} {\bibinfo {author} {\bibfnamefont {Simon}\ \bibnamefont
  {Capstick}}\ and\ \bibinfo {author} {\bibfnamefont {W.}~\bibnamefont
  {Roberts}},\ }\bibfield  {title} {\enquote {\bibinfo {title} {{Quark models
  of baryon masses and decays}},}\ }\href {\doibase
  10.1016/S0146-6410(00)00109-5} {\bibfield  {journal} {\bibinfo  {journal}
  {Prog. Part. Nucl. Phys.}\ }\textbf {\bibinfo {volume} {45}},\ \bibinfo
  {pages} {S241--S331} (\bibinfo {year} {2000})},\ \Eprint
  {http://arxiv.org/abs/nucl-th/0008028} {arXiv:nucl-th/0008028} \BibitemShut
  {NoStop}%
\bibitem [{\citenamefont {Gross}\ \emph {et~al.}(2023)\citenamefont {Gross}
  \emph {et~al.}}]{Gross:2022hyw}%
  \BibitemOpen
  \bibfield  {author} {\bibinfo {author} {\bibfnamefont {Franz}\ \bibnamefont
  {Gross}} \emph {et~al.},\ }\bibfield  {title} {\enquote {\bibinfo {title}
  {{50 Years of Quantum Chromodynamics}},}\ }\href {\doibase
  10.1140/epjc/s10052-023-11949-2} {\bibfield  {journal} {\bibinfo  {journal}
  {Eur. Phys. J. C}\ }\textbf {\bibinfo {volume} {83}},\ \bibinfo {pages}
  {1125} (\bibinfo {year} {2023})},\ \Eprint {http://arxiv.org/abs/2212.11107}
  {arXiv:2212.11107 [hep-ph]} \BibitemShut {NoStop}%
\bibitem [{\citenamefont {Shuryak}\ and\ \citenamefont
  {Vainshtein}(1982)}]{Shuryak:1981kj}%
  \BibitemOpen
  \bibfield  {author} {\bibinfo {author} {\bibfnamefont {Edward~V.}\
  \bibnamefont {Shuryak}}\ and\ \bibinfo {author} {\bibfnamefont {A.~I.}\
  \bibnamefont {Vainshtein}},\ }\bibfield  {title} {\enquote {\bibinfo {title}
  {{Theory of Power Corrections to Deep Inelastic Scattering in Quantum
  Chromodynamics. 1. Q**2 Effects}},}\ }\href {\doibase
  10.1016/0550-3213(82)90355-8} {\bibfield  {journal} {\bibinfo  {journal}
  {Nucl. Phys. B}\ }\textbf {\bibinfo {volume} {199}},\ \bibinfo {pages}
  {451--481} (\bibinfo {year} {1982})}\BibitemShut {NoStop}%
\bibitem [{\citenamefont {Shuryak}\ and\ \citenamefont
  {Zahed}(2023{\natexlab{a}})}]{Shuryak:2021fsu}%
  \BibitemOpen
  \bibfield  {author} {\bibinfo {author} {\bibfnamefont {Edward}\ \bibnamefont
  {Shuryak}}\ and\ \bibinfo {author} {\bibfnamefont {Ismail}\ \bibnamefont
  {Zahed}},\ }\bibfield  {title} {\enquote {\bibinfo {title} {{Hadronic
  structure on the light front. I. Instanton effects and quark-antiquark
  effective potentials}},}\ }\href {\doibase 10.1103/PhysRevD.107.034023}
  {\bibfield  {journal} {\bibinfo  {journal} {Phys. Rev. D}\ }\textbf {\bibinfo
  {volume} {107}},\ \bibinfo {pages} {034023} (\bibinfo {year}
  {2023}{\natexlab{a}})},\ \Eprint {http://arxiv.org/abs/2110.15927}
  {arXiv:2110.15927 [hep-ph]} \BibitemShut {NoStop}%
\bibitem [{\citenamefont {Puckett}\ \emph {et~al.}(2012)\citenamefont {Puckett}
  \emph {et~al.}}]{Puckett:2011xg}%
  \BibitemOpen
  \bibfield  {author} {\bibinfo {author} {\bibfnamefont {A.~J.~R.}\
  \bibnamefont {Puckett}} \emph {et~al.},\ }\bibfield  {title} {\enquote
  {\bibinfo {title} {{Final Analysis of Proton Form Factor Ratio Data at
  $\mathbf{Q^2 = 4.0}$, 4.8 and 5.6 GeV$\mathbf{^2}$}},}\ }\href {\doibase
  10.1103/PhysRevC.85.045203} {\bibfield  {journal} {\bibinfo  {journal} {Phys.
  Rev. C}\ }\textbf {\bibinfo {volume} {85}},\ \bibinfo {pages} {045203}
  (\bibinfo {year} {2012})},\ \Eprint {http://arxiv.org/abs/1102.5737}
  {arXiv:1102.5737 [nucl-ex]} \BibitemShut {NoStop}%
\bibitem [{\citenamefont {Shuryak}\ and\ \citenamefont
  {Zahed}(2021)}]{Shuryak:2020ktq}%
  \BibitemOpen
  \bibfield  {author} {\bibinfo {author} {\bibfnamefont {Edward}\ \bibnamefont
  {Shuryak}}\ and\ \bibinfo {author} {\bibfnamefont {Ismail}\ \bibnamefont
  {Zahed}},\ }\bibfield  {title} {\enquote {\bibinfo {title} {{Nonperturbative
  quark-antiquark interactions in mesonic form factors}},}\ }\href {\doibase
  10.1103/PhysRevD.103.054028} {\bibfield  {journal} {\bibinfo  {journal}
  {Phys. Rev. D}\ }\textbf {\bibinfo {volume} {103}},\ \bibinfo {pages}
  {054028} (\bibinfo {year} {2021})},\ \Eprint
  {http://arxiv.org/abs/2008.06169} {arXiv:2008.06169 [hep-ph]} \BibitemShut
  {NoStop}%
\bibitem [{\citenamefont {Simonov}(2021)}]{Simonov:2020wql}%
  \BibitemOpen
  \bibfield  {author} {\bibinfo {author} {\bibfnamefont {Yu.~A.}\ \bibnamefont
  {Simonov}},\ }\bibfield  {title} {\enquote {\bibinfo {title} {{Proton and
  neutron form factors with quark orbital excitations}},}\ }\href {\doibase
  10.1140/epja/s10050-021-00546-0} {\bibfield  {journal} {\bibinfo  {journal}
  {Eur. Phys. J. A}\ }\textbf {\bibinfo {volume} {57}},\ \bibinfo {pages} {228}
  (\bibinfo {year} {2021})},\ \Eprint {http://arxiv.org/abs/2010.12666}
  {arXiv:2010.12666 [hep-ph]} \BibitemShut {NoStop}%
\bibitem [{\citenamefont {Shuryak}\ and\ \citenamefont
  {Zahed}(2023{\natexlab{b}})}]{Shuryak:2022wtk}%
  \BibitemOpen
  \bibfield  {author} {\bibinfo {author} {\bibfnamefont {Edward}\ \bibnamefont
  {Shuryak}}\ and\ \bibinfo {author} {\bibfnamefont {Ismail}\ \bibnamefont
  {Zahed}},\ }\bibfield  {title} {\enquote {\bibinfo {title} {{Hadronic
  structure on the light front. V. Diquarks, nucleons, and multiquark Fock
  components}},}\ }\href {\doibase 10.1103/PhysRevD.107.034027} {\bibfield
  {journal} {\bibinfo  {journal} {Phys. Rev. D}\ }\textbf {\bibinfo {volume}
  {107}},\ \bibinfo {pages} {034027} (\bibinfo {year} {2023}{\natexlab{b}})},\
  \Eprint {http://arxiv.org/abs/2208.04428} {arXiv:2208.04428 [hep-ph]}
  \BibitemShut {NoStop}%
\bibitem [{\citenamefont {Deur}\ \emph {et~al.}(2018)\citenamefont {Deur},
  \citenamefont {Brodsky},\ and\ \citenamefont {De~T\'eramond}}]{Deur:2018roz}%
  \BibitemOpen
  \bibfield  {author} {\bibinfo {author} {\bibfnamefont {Alexandre}\
  \bibnamefont {Deur}}, \bibinfo {author} {\bibfnamefont {Stanley~J.}\
  \bibnamefont {Brodsky}}, \ and\ \bibinfo {author} {\bibfnamefont {Guy~F.}\
  \bibnamefont {De~T\'eramond}},\ }\bibfield  {title} {\enquote {\bibinfo
  {title} {{The Spin Structure of the Nucleon}},}\ }\href {\doibase
  10.1088/1361-6633/ab0b8f} {\  (\bibinfo {year} {2018}),\
  10.1088/1361-6633/ab0b8f},\ \Eprint {http://arxiv.org/abs/1807.05250}
  {arXiv:1807.05250 [hep-ph]} \BibitemShut {NoStop}%
\bibitem [{\citenamefont {Liu}\ \emph {et~al.}(2024)\citenamefont {Liu},
  \citenamefont {Shuryak},\ and\ \citenamefont {Zahed}}]{Liu:2024rdm}%
  \BibitemOpen
  \bibfield  {author} {\bibinfo {author} {\bibfnamefont {Wei-Yang}\
  \bibnamefont {Liu}}, \bibinfo {author} {\bibfnamefont {Edward}\ \bibnamefont
  {Shuryak}}, \ and\ \bibinfo {author} {\bibfnamefont {Ismail}\ \bibnamefont
  {Zahed}},\ }\bibfield  {title} {\enquote {\bibinfo {title} {{Glue in hadrons
  at medium resolution and the QCD instanton vacuum}},}\ }\href@noop {} {\
  (\bibinfo {year} {2024})},\ \Eprint {http://arxiv.org/abs/2404.03047}
  {arXiv:2404.03047 [hep-ph]} \BibitemShut {NoStop}%
\bibitem [{\citenamefont {Shuryak}(2021)}]{Shuryak:2021vnj}%
  \BibitemOpen
  \bibfield  {author} {\bibinfo {author} {\bibfnamefont {Edward}\ \bibnamefont
  {Shuryak}},\ }\href {\doibase 10.1007/978-3-030-62990-8} {\emph {\bibinfo
  {title} {{Nonperturbative Topological Phenomena in QCD and Related
  Theories}}}},\ \bibinfo {series} {Lecture Notes in Physics}, Vol.\ \bibinfo
  {volume} {977}\ (\bibinfo {year} {2021})\BibitemShut {NoStop}%
\bibitem [{\citenamefont {Eichten}\ and\ \citenamefont
  {Feinberg}(1981)}]{Eichten:1980mw}%
  \BibitemOpen
  \bibfield  {author} {\bibinfo {author} {\bibfnamefont {E.}~\bibnamefont
  {Eichten}}\ and\ \bibinfo {author} {\bibfnamefont {F.}~\bibnamefont
  {Feinberg}},\ }\bibfield  {title} {\enquote {\bibinfo {title} {{Spin
  Dependent Forces in QCD}},}\ }\href {\doibase 10.1103/PhysRevD.23.2724}
  {\bibfield  {journal} {\bibinfo  {journal} {Phys. Rev. D}\ }\textbf {\bibinfo
  {volume} {23}},\ \bibinfo {pages} {2724} (\bibinfo {year}
  {1981})}\BibitemShut {NoStop}%
\bibitem [{\citenamefont {Miesch}\ \emph {et~al.}(2025)\citenamefont {Miesch},
  \citenamefont {Shuryak},\ and\ \citenamefont {Zahed}}]{Miesch:2024fhv}%
  \BibitemOpen
  \bibfield  {author} {\bibinfo {author} {\bibfnamefont {Nicholas}\
  \bibnamefont {Miesch}}, \bibinfo {author} {\bibfnamefont {Edward}\
  \bibnamefont {Shuryak}}, \ and\ \bibinfo {author} {\bibfnamefont {Ismail}\
  \bibnamefont {Zahed}},\ }\bibfield  {title} {\enquote {\bibinfo {title}
  {{Bridging hadronic and vacuum structure by heavy quarkonia}},}\ }\href
  {\doibase 10.1103/PhysRevD.111.034006} {\bibfield  {journal} {\bibinfo
  {journal} {Phys. Rev. D}\ }\textbf {\bibinfo {volume} {111}},\ \bibinfo
  {pages} {034006} (\bibinfo {year} {2025})},\ \Eprint
  {http://arxiv.org/abs/2403.18700} {arXiv:2403.18700 [hep-ph]} \BibitemShut
  {NoStop}%
\bibitem [{\citenamefont {Strominger}(1981)}]{Strominger:1980xa}%
  \BibitemOpen
  \bibfield  {author} {\bibinfo {author} {\bibfnamefont {Andrew}\ \bibnamefont
  {Strominger}},\ }\bibfield  {title} {\enquote {\bibinfo {title} {{Loop Space
  Solution of Two-dimensional {QCD}}},}\ }\href {\doibase
  10.1016/0370-2693(81)90311-7} {\bibfield  {journal} {\bibinfo  {journal}
  {Phys. Lett. B}\ }\textbf {\bibinfo {volume} {101}},\ \bibinfo {pages}
  {271--276} (\bibinfo {year} {1981})}\BibitemShut {NoStop}%
\bibitem [{\citenamefont {Polyakov}(1988)}]{Polyakov:1988md}%
  \BibitemOpen
  \bibfield  {author} {\bibinfo {author} {\bibfnamefont {Alexander~M.}\
  \bibnamefont {Polyakov}},\ }\bibfield  {title} {\enquote {\bibinfo {title}
  {{Fermi-Bose Transmutations Induced by Gauge Fields}},}\ }\href {\doibase
  10.1142/S0217732388000398} {\bibfield  {journal} {\bibinfo  {journal} {Mod.
  Phys. Lett. A}\ }\textbf {\bibinfo {volume} {3}},\ \bibinfo {pages} {325}
  (\bibinfo {year} {1988})}\BibitemShut {NoStop}%
\bibitem [{\citenamefont {Kogut}\ and\ \citenamefont
  {Parisi}(1981)}]{Kogut:1981gm}%
  \BibitemOpen
  \bibfield  {author} {\bibinfo {author} {\bibfnamefont {John~B.}\ \bibnamefont
  {Kogut}}\ and\ \bibinfo {author} {\bibfnamefont {G.}~\bibnamefont {Parisi}},\
  }\bibfield  {title} {\enquote {\bibinfo {title} {{Long Range Spin Spin Forces
  in Gauge Theories}},}\ }\href {\doibase 10.1103/PhysRevLett.47.1089}
  {\bibfield  {journal} {\bibinfo  {journal} {Phys. Rev. Lett.}\ }\textbf
  {\bibinfo {volume} {47}},\ \bibinfo {pages} {1089} (\bibinfo {year}
  {1981})}\BibitemShut {NoStop}%
\bibitem [{\citenamefont {Burkert}\ and\ \citenamefont
  {Roberts}(2019)}]{Burkert:2017djo}%
  \BibitemOpen
  \bibfield  {author} {\bibinfo {author} {\bibfnamefont {Volker~D.}\
  \bibnamefont {Burkert}}\ and\ \bibinfo {author} {\bibfnamefont {Craig~D.}\
  \bibnamefont {Roberts}},\ }\bibfield  {title} {\enquote {\bibinfo {title}
  {{Colloquium : Roper resonance: Toward a solution to the fifty year
  puzzle}},}\ }\href {\doibase 10.1103/RevModPhys.91.011003} {\bibfield
  {journal} {\bibinfo  {journal} {Rev. Mod. Phys.}\ }\textbf {\bibinfo {volume}
  {91}},\ \bibinfo {pages} {011003} (\bibinfo {year} {2019})},\ \Eprint
  {http://arxiv.org/abs/1710.02549} {arXiv:1710.02549 [nucl-ex]} \BibitemShut
  {NoStop}%
\bibitem [{\citenamefont {Brown}\ and\ \citenamefont
  {Bolsterli}(1959)}]{Brown:1959zzb}%
  \BibitemOpen
  \bibfield  {author} {\bibinfo {author} {\bibfnamefont {G.~E.}\ \bibnamefont
  {Brown}}\ and\ \bibinfo {author} {\bibfnamefont {M.}~\bibnamefont
  {Bolsterli}},\ }\bibfield  {title} {\enquote {\bibinfo {title} {{Dipole State
  in Nuclei}},}\ }\href {\doibase 10.1103/PhysRevLett.3.472} {\bibfield
  {journal} {\bibinfo  {journal} {Phys. Rev. Lett.}\ }\textbf {\bibinfo
  {volume} {3}},\ \bibinfo {pages} {472--476} (\bibinfo {year}
  {1959})}\BibitemShut {NoStop}%
\bibitem [{\citenamefont {Zahed}\ \emph {et~al.}(1984)\citenamefont {Zahed},
  \citenamefont {Meissner},\ and\ \citenamefont {Kaulfuss}}]{Zahed:1984qv}%
  \BibitemOpen
  \bibfield  {author} {\bibinfo {author} {\bibfnamefont {I.}~\bibnamefont
  {Zahed}}, \bibinfo {author} {\bibfnamefont {U.~G.}\ \bibnamefont {Meissner}},
  \ and\ \bibinfo {author} {\bibfnamefont {U.~B.}\ \bibnamefont {Kaulfuss}},\
  }\bibfield  {title} {\enquote {\bibinfo {title} {{LOW LYING RESONANCES IN THE
  SKYRME MODEL USING THE SEMICLASSICAL APPROXIMATION}},}\ }\href {\doibase
  10.1016/0375-9474(84)90162-3} {\bibfield  {journal} {\bibinfo  {journal}
  {Nucl. Phys. A}\ }\textbf {\bibinfo {volume} {426}},\ \bibinfo {pages}
  {525--541} (\bibinfo {year} {1984})}\BibitemShut {NoStop}%
\bibitem [{\citenamefont {He}\ and\ \citenamefont {Zahed}(2024)}]{He:2024jgc}%
  \BibitemOpen
  \bibfield  {author} {\bibinfo {author} {\bibfnamefont {Fangcheng}\
  \bibnamefont {He}}\ and\ \bibinfo {author} {\bibfnamefont {Ismail}\
  \bibnamefont {Zahed}},\ }\bibfield  {title} {\enquote {\bibinfo {title}
  {{Helium-4 gravitational form factors: Exchange currents}},}\ }\href
  {\doibase 10.1103/PhysRevC.110.054302} {\bibfield  {journal} {\bibinfo
  {journal} {Phys. Rev. C}\ }\textbf {\bibinfo {volume} {110}},\ \bibinfo
  {pages} {054302} (\bibinfo {year} {2024})},\ \Eprint
  {http://arxiv.org/abs/2406.07412} {arXiv:2406.07412 [nucl-th]} \BibitemShut
  {NoStop}%
\bibitem [{\citenamefont {Gallimore}\ and\ \citenamefont
  {Liao}(2024)}]{Gallimore:2024fcz}%
  \BibitemOpen
  \bibfield  {author} {\bibinfo {author} {\bibfnamefont {Daniel}\ \bibnamefont
  {Gallimore}}\ and\ \bibinfo {author} {\bibfnamefont {Jinfeng}\ \bibnamefont
  {Liao}},\ }\bibfield  {title} {\enquote {\bibinfo {title} {{A Potential Model
  Study of the Nucleon's Charge and Mass Radius}},}\ }\href@noop {} {\
  (\bibinfo {year} {2024})},\ \Eprint {http://arxiv.org/abs/2405.07077}
  {arXiv:2405.07077 [hep-ph]} \BibitemShut {NoStop}%
\end{thebibliography}%
\end{document}